\documentclass[preprint2,numberedappendix]{emulateapj-rtx4}
\usepackage{graphicx,graphics,amsmath}
\usepackage{graphicx,natbib}
\usepackage{bm,url}
\graphicspath{{./fig/}{./png/}}
\newcommand{\EQ}{\begin{equation}}
\newcommand{\EN}{\end{equation}}
\newcommand{\EQA}{\begin{eqnarray}}
\newcommand{\ENA}{\end{eqnarray}}
\newcommand{\nab}{\mbox{\boldmath $\nabla$} {}}

\newcommand{\DIV}{\bm{\nabla} \cdot }

\newcommand{\AAA}{\bm{{A}}}
\newcommand{\BB}{\bm{{B}}}
\newcommand{\JJ}{\bm{{J}}}
\newcommand{\UU}{\bm{{U}}}

\newcommand{\aaaa}{\mbox{\boldmath $a$} {}}

\newcommand{\meanA}{\overline{A}}
\newcommand{\meanB}{\overline{B}}

\newcommand{\meanJ}{\overline{J}}

\newcommand{\meanv}[1]{\overline{\bm #1}}

\newcommand{\meanemf}{\overline{\cal E} {}}
\newcommand{\meanEMF}{\overline{\mbox{\boldmath ${\mathcal E}$}} {}}

\newcommand{\meanBB}{\bm{\overline{{B}}}}
\newcommand{\meanJJ}{\bm{\overline{{J}}}}
\newcommand{\meanKK}{\bm{\overline{{K}}}}
\newcommand{\meanUU}{\bm{\overline{{U}}}}
\newcommand{\oo}{\bm{{{\omega}}}}
\newcommand{\uu}{\bm{{{u}}}}
\newcommand{\bb}{\bm{{{b}}}}
\newcommand{\ee}{\bm{{{e}}}}
\newcommand{\jj}{\bm{{{j}}}}
\newcommand{\ff}{\bm{{{f}}}}
\newcommand{\kk}{\bm{{{k}}}}
\newcommand{\xx}{\bm{{{x}}}}

\newcommand{\eee}{\hat{\mbox{\boldmath $e$}} {}}
\newcommand{\xxx}{\hat{\mbox{\boldmath $x$}} {}}
\newcommand{\zzz}{\hat{\bm{z}}}
\newcommand{\zz}{\hat{\mbox{\boldmath $z$}} {}}
\newcommand{\s}{\,{\rm s}}
\newcommand{\cm}{\,{\rm cm}}
\newcommand{\yr}{\,{\rm yr}}
\newcommand{\aaT}{\aaaa^{\rm T}}
\newcommand{\corr}{{\,\rm corr}}

\def\kf{k_\mathrm{f}}
\def\ii{{\rm i}}
\def\dd{{\rm d}}

\def\Ta{\mbox{\rm Ta}}
\def\Ra{\mbox{\rm Ra}}

\def\Pra{\mbox{\rm Pr}}

\def\Pm{P_\mathrm{m}}
\def\Rm{R_\mathrm{m}}
\def\Rmz{R_\mathrm{m0}}
\def\Rey{\mbox{\rm Re}}

\newcommand{\etat}{\eta_{\rm t}}
\newcommand{\etaT}{\eta_{\rm T}}

\newcommand{\etatz}{\eta_{\rm t0}}

\newcommand{\cs}{c_{\rm s}}
\newcommand{\urms}{u_{\rm rms}}
\newcommand{\urmsh}{u_{\rm rms0}}

\newcommand{\Beq}{B_{\rm eq}}
\newcommand{\Beqz}{B_{\rm eq0}}
\newcommand{\Beqh}{B_{\rm eq0}}
\newcommand{\Bex}{B_{\rm ext}}
\newcommand{\BBex}{\bm{B}_{\rm ext}}
\newcommand{\hatmeanBB}{\hat{\meanBB}}
\newcommand{\hatmeanB}{\hat{\meanB}}

\newcommand{\xu}{\hat{\bm x}}

\newcommand{\zu}{\hat{\bm z}}

\newcommand{\kef}{k_{\rm f}}

\def\onethird{{\textstyle{1\over3}}}
\def\onehalf{{\textstyle{1\over2}}}

\def\onethird{{\textstyle{1\over3}}}

\newcommand{\sz}{\,{\rm s}z}
\newcommand{\cx}{\,{\rm c}x}
\newcommand{\cy}{\,{\rm c}y}
\newcommand{\cz}{\,{\rm c}z}

\newcommand{\SSSS}{\mbox{\boldmath ${\sf S}$} {}}

\newcommand{\Eq}[1]{Equation~(\ref{#1})}
\newcommand{\Eqs}[2]{equations~(\ref{#1}) and~(\ref{#2})}

\newcommand{\App}[1]{Appendix~\ref{#1}}
\newcommand{\Sec}[1]{\S\ref{#1}}

\newcommand{\Fig}[1]{Figure~\ref{#1}}
\newcommand{\Tab}[1]{Table~\ref{#1}}

\newcommand{\bra}[1]{\langle #1\rangle}
\def\tfz{TFZ}
\def\tfa{TFA}
\begin{document}

\title{Quenching and anisotropy of hydromagnetic turbulent transport}
\author{Bidya Binay Karak$^{1}$, Matthias Rheinhardt$^{2}$, Axel Brandenburg$^{1,3}$, Petri J. K\"apyl\"a$^{2,4}$, and Maarit J. K\"apyl\"a$^{4}$}
\affil{$^1$Nordita, KTH Royal Institute of Technology and Stockholm University, Roslagstullsbacken 23, SE-10691 Stockholm, Sweden\\
$^2$Department of Physics, Gustaf H\"allstr\"omin katu 2a (PO Box 64), FI-00014 University of Helsinki, Finland\\
$^3$Department of Astronomy, AlbaNova University Center, Stockholm University, SE-10691 Stockholm, Sweden\\
$^4$ReSoLVE Centre of Excellence, Dept of Information and Computer Science, Aalto University, PO Box 15400, FI-00076 Aalto, Finland\\
}

\date{\small\today,~ $ $Revision: 1.378 $ $}

\begin{abstract}
Hydromagnetic turbulence affects the evolution of large-scale
magnetic fields through mean-field effects like turbulent diffusion 
and the $\alpha$ effect.
For stronger fields, these effects are usually suppressed or quenched,
and additional anisotropies are introduced.
Using different variants of the test-field method,
we determine the quenching of the turbulent transport coefficients for
the forced Roberts flow, isotropically forced non-helical 
turbulence, and rotating thermal convection.
We see significant quenching only when
the mean magnetic field is larger than the equipartition value of the turbulence.
Expressing the magnetic field in terms of the equipartition value
of the {\it quenched} flows, we obtain for the quenching exponents
of the turbulent magnetic diffusivity about 1.3, 1.1, and 1.3 
for Roberts flow, forced turbulence, and convection, respectively.
However, when the magnetic field is expressed in terms of the equipartition value
of the {\it unquenched} flows these quenching exponents
become about 4, 1.5, and 2.3, respectively.
For the $\alpha$ effect, the exponent is about 1.3 
for the Roberts flow and 2 for convection in the first case,
but 4 and 3, respectively, in the second. 
In convection, the quenching of turbulent pumping follows 
the same power law as turbulent diffusion, while for
the coefficient describing the $\bm\varOmega\times\JJ$ effect 
nearly the same quenching exponent is obtained as for $\alpha$.
For forced turbulence,
turbulent diffusion proportional to the second derivative 
along the mean magnetic field is quenched much less, especially for
larger values of the magnetic Reynolds number.
However, we find that
in corresponding axisymmetric mean-field dynamos with dominant toroidal 
field the quenched diffusion coefficients
are the same for the poloidal and toroidal
field constituents.
\end{abstract}

\section{Introduction}

Many astrophysical objects possess turbulent convection, and the dynamo
mechanisms based on it are believed to be responsible for the generation and
maintenance of the observed magnetic fields. 
The study of the dynamo mechanism in the solar convection zone using
simulations of turbulent convection in spherical shells began 
in the 1980s with the works of \cite{GM81}, \cite{Gil83} and
\cite{Gla85}, and has recently been
pursued further by many more authors \citep{brun04, racine11, K12, K13, KKKB14}.
However, under stellar conditions the dimensionless parameters governing
magnetohydrodynamics attain extreme values,
which are far from being accessible through numerical models.
So we do not know to what extent feasible models at 
temperate parameter regimes reflect properties 
of convection and dynamos in real stars.
An alternative approach to studying the dynamo problem
is mean-field theory which began with
the pioneering works of \cite{Par55}, \cite{Bra64}, and \cite{SKR66}.
This approach is computationally less expensive because one needs not
to resolve the full dynamical range of the small-scale turbulence,
which is instead parameterized. In recent years, there have been significant achievements
of mean-field MHD in reproducing various aspects of magnetic and flow fields
in the Sun \citep[e.g.,][]{CNC04,rempel06,KKT06,CK09,CK12,Ka10,char10,PK11}.

In this context, an important task is to determine the mean electromotive
force $\meanEMF$, which results from the correlation between the fluctuating
constituents of velocity and magnetic field, in terms of the mean field $\meanBB$.
There is no accurate theory to accomplish this task from first principles, 
except for some limiting cases, in particular those of small
Strouhal and magnetic Reynolds number, $\Rm$.
Therefore, suitable assumptions are required in determining $\meanEMF$.
When $\meanBB$ varies slowly in space and time, we may write
\EQ
\meanemf_i=\alpha_{ij}\meanB_{j} + \beta_{ijk}\frac{\partial \meanB_{j}}{\partial  x_k}.
\label{meanemf1}
\EN
The diagonal components of $\alpha_{ij}$ are usually the most
important terms for dynamo action, but
in the presence of shear, the $\bm\varOmega\times\JJ$
\citep{KR80} and shear-current \citep{RK03} effects,
both covered by $\beta_{ijk}$, can also enable it.
Many components of $\beta_{ijk}$, however, describe dissipative effects.

Doubts can be raised regarding the explanatory and predictive power
of mean-field dynamo models given that
the tensors $\alpha_{ij}$ and $\beta_{ijk}$ are often chosen to some extent arbitrarily 
or even tuned to obtain results resembling features of the Sun. 
Therefore, methods to measure these coefficients from simulations have been developed. 
At present the most accurate method is the so-called 
test-field method \citep{Sch05,Sch07,BRRS08,BGKKMR13}. In this method, 
one selects an adequate number of independent
mean fields, the
`test fields',
and solves for each of them the corresponding equation for the
fluctuating magnetic field (in addition to the main simulation).
Finally, via computing the mean electromotive force,
the transport coefficients are calculated. 

There are different variants of the test-field method.
The best established one is
based on the average over two spatial 
(the `horizontal')
coordinates.
This method has been applied to a large variety of 
setups, e.g., isotropic homogeneous turbulence \citep{SBS08,BRRS08},
homogeneous shear flow turbulence \citep{BRRK08}, 
with and without helicity \citep{MKTB09}, turbulent convection \citep{KKB09a}, 
and supernova driven interstellar turbulence \citep{G13}.
Another variant is based on Fourier-weighted horizontal averages
and allows to determine also the coefficients that multiply  
horizontal derivatives of the mean field.
This method has been applied to forced turbulence \citep{BRK12} and
to cosmic ray-driven turbulence \citep{RKBE12,BBMO13}.

In models based on thin flux tubes, 
forming the major alternative to distributed turbulent dynamos,
the magnetic field strength
in the deep parts of the solar convection zone is believed to exceed 
its value at equipartition with velocity \citep{CG87,DC93,WFM11}.
On the other hand, it is well known that turbulent transport becomes less efficient when
the mean magnetic field's strength is comparable to or larger
than the equipartition value. Therefore precise
knowledge of the quenching is
 needed.
Mean-field dynamo models of the $\alpha\varOmega$ 
type
often employ an `ad hoc' algebraic or 
dynamical $\alpha$-quenching \citep{Jep75, covas98},
while largely ignoring the quenching of 
the turbulent diffusivity
$\etat$ despite of its
importance in determining the cycle frequency.
Indeed, in the absence of quenching, the standard estimate of $\etat$
for the Sun ($\sim 10^{12}$--$10^{13}\cm^2\s^{-1}$) yields a rather
short cycle period of $2$--$3\yr$ \citep{Koh73}.
However, by considering the quenching of $\etat$, a reasonable value
of the cycle period can easily be obtained \citep{R94,GDD09,MNM11}.
In fact, measuring the cycle frequency in a simulation has been one way
of determining the quenching of $\etat$ \citep{KB09}.

Early work by \cite{Mof72} and \cite{Rue74} showed that under the
Second Order Correlation Approximation (SOCA), $\alpha$ is quenched
inversely proportional to the third power of the magnetic field. 
Following \cite{VC92}, several investigations have suggested that $\alpha$ is
beginning to be quenched noticeably when the mean field becomes comparable to
$\Rm^{-1}$ times the equipartition value \citep{CH96},
i.e., for extremely weak magnetic fields.
This behavior is also called `catastrophic quenching'.
However, it is now understood as an artifact of having defined
volume-averaged mean fields \citep{B01,BRRS08} combined with the usage
of perfectly conducting
or periodic boundary conditions and is not expected to be 
important in astrophysical bodies where magnetic helicity fluxes
can alleviate catastrophic quenching \citep[e.g.,][]{KMRS00,DSGB13}.
The actual value of $\alpha$ shows a much
weaker dependence on $\Rm$ even when $\meanB$ is comparable with
the equipartition value \citep{BRRS08}.
This work also shows that the $\Rm$ dependence of
$\alpha$ and  $\etat$ is such that their contributions
to the growth rate nearly balance, with a residual matching the microscopic
resistive term. 
This is in fact a requirement for the dynamo to be in a saturated state.
Consequently the saturated
mean electromotive force is proportional to $\Rm^{-1}$,
which is sometimes misinterpreted as catastrophic quenching.

Once catastrophic quenching is alleviated, the magnetic field can grow
to equipartition field strengths, 
when other quenching mechanisms
that are not $\Rm$ dependent might become important and can therefore
be studied already for smaller values of $\Rm$.
\cite{SBS07} found that $\alpha$ is quenched proportional to $1/\meanB^2$
and $1/\meanB^3$ for time-dependent and steady flows, respectively.
Their latter result was based on analytic theory and appeared to be confirmed
by numerical simulations using a 
steady forcing
proportional to the ABC-flow.
However, subsequent work by \cite{RB10} demonstrated quenching proportional
to $\meanB^{\,-4}$ for a steady forcing proportional to the flow I of
\cite{Rob72}, hereafter referred to as Roberts flow.
They also noted that for 
ABC-flow forcing
the quenching is indeed
better described 
when setting the power also to 4 instead of 3.
More recently, in supernova driven 
 turbulent dynamo simulations, \citet{G13} find
$\alpha \sim (\meanB/\Beq)^{-2}$ where $\Beq$ is the local equipartition value.

For the turbulent diffusivity,
\cite{KPR94} and \cite{RK00} obtained that $\etat$ is quenched
inversely proportional to $\meanB$.
In the two-dimensional case, \cite{CV91} have found catastrophic quenching
of $\etat$. However, this is a special 
situation
connected with the
fact that in two dimensions the mean square vector potential
is a conserved quantity. This is no longer the case in three dimensions.
Quenching similar to \cite{KPR94} has been confirmed by simulations \citep{B01,BB02,G13}.
In particular, making the ansatz
$\etat \sim 1/\big(1 + p (\meanB/\Beq)^{q}\big)$, 
\cite{G13} find
$q \approx 1$ in supernova-driven simulations of the turbulent interstellar medium.
On the other hand, \cite{YBR03} find $q \approx 2$ in
simulations of forced turbulence with a decaying large-scale magnetic field.
However \cite{KB09} found that their results depend on the
strength of shear and found $q \approx 1$ for weak shear while
$q \approx 2$ for strong shear.

In the present work we measure the quenching of these transport
coefficients as a function of the mean magnetic field strength for
three different background simulations --
(i) a forced Roberts flow,
(ii) forced turbulence in a triply periodic box, and
(iii) convection in a bounded box.
In all these simulations we impose a uniform and constant
external mean field.
However, this induces a preferred
direction which causes the {\em statistical properties} of the turbulence to
be axisymmetric with respect to the 
direction of the
magnetic field.
In the following, we refer to such flows as {\em axisymmetric turbulence},
for which the number of independent components of the
$\alpha$ and $\eta$ tensors is reduced to only nine, 
simplifying also their determination \citep{BRK12}.

\section{Concept of turbulent transport in mean-field dynamo}
\label{mfconcept}

The evolution of the magnetic field $\BB$ in an electrically
conducting fluid is governed by the induction equation
\EQ
{\partial\BB\over\partial t}=\nab\times\left(\UU\times\BB-\eta\JJ\right),
\label{indeq}
\EN
where $\UU$ is the fluid velocity. Here, $\eta$ is the microphysical magnetic diffusivity,
while the magnetic permeability of the fluid has been set to unity. Thus the current density 
$\JJ$ is given by $\JJ = \nab\times\BB$. In mean-field MHD, we consider the fields as
sums of `averaged' and small-scale `fluctuating' fields,
with the assumption that the averaging satisfies 
(at least approximately) the Reynolds rules.
Denoting averaged fields by overbars and fluctuating ones by lowercase letters, 
we write the equation for the mean magnetic field $\meanBB$ as
\EQ
{\partial\meanBB\over\partial t}=\nab\times\left(\meanUU\times\meanBB+\meanEMF-\eta\meanJJ\right),
\label{mean_indeq}
\EN
where $\meanEMF=\overline{\uu\times\bb}$ is the aforementioned mean electromotive force,
which captures the correlation of the fluctuating fields $\uu$ and $\bb$.
The ultimate goal of mean-field MHD is to express $\meanEMF$
in terms of $\meanBB$ itself. There are several procedures
for doing that. When the mean magnetic field varies slowly in space and time
we can write $\meanEMF$ in the form of Equation~(\ref{meanemf1}).
Our primary goal is to measure the transport coefficients $\alpha_{ij}$ and $\eta_{ijk}$ in
the presence of an imposed  uniform magnetic field $\BB_{\rm ext}$ and, in particular,
measure the degree of their quenching and anisotropy.

Let us consider turbulence that is anisotropic and exhibiting
only one preferred direction $\eee$, referring to an external magnetic field, rotation axis, or the direction of gravity. 
Then following \cite{BRK12}, the general
representation of $\meanEMF$ is given by
\begin{alignat}{5}
\meanEMF =\phantom{-}&
\alpha_\perp\meanBB
&&+(\alpha_\parallel-\alpha_\perp)&&(\eee\cdot\meanBB)&&\eee
&&+\gamma\eee\times\meanBB
\nonumber \\
-  &\eta_\perp\meanJJ
&&-(\eta_\parallel-\eta_\perp)&&(\eee\cdot\meanJJ)&&\eee
&&-\delta\eee\times\meanJJ
\label{axisymE}\\
-&\kappa_\perp\meanKK
&&-(\kappa_\parallel-\kappa_\perp)&&(\eee\cdot\meanKK)&&\eee
&&-\mu\eee\times\meanKK
\nonumber
\end{alignat}
with nine coefficients $\alpha_\perp$, $\alpha_\parallel$, $\ldots$, $\mu$.
While $\meanJJ$ is given by the antisymmetric part
of the gradient tensor $\nab \meanBB$, $\meanKK$ is 
defined by $\meanKK=\eee \cdot (\nab\meanBB)^\mathrm{S}$
with $(\nab\meanBB)^\mathrm{S}$ being the symmetric part of $\nab \meanBB$.
For homogeneous isotropic turbulence $\alpha_\parallel=\alpha_\perp$, 
$\eta_\parallel=\eta_\perp$ and the other coefficients vanish.
We note that our sign convention for $\alpha_\perp$, $\alpha_\parallel$,
and $\gamma$ follows that commonly used, but it differs from that used
in \cite{BRK12}.

The $\mu$ term corresponds to a modification of turbulent diffusion
along the preferred direction.
To understand this, let us assume 
that only $\eta_\perp$, $\eta_\|$, and $\mu$ are non-vanishing
and independent of position.
By introducing the quantities 
$\etaT\equiv\etat+\eta$ with $\etat\equiv\eta_\perp-\mu/2$ 
and $\epsilon\equiv\eta_\parallel-\eta_\perp+\mu/2$, we have
\begin{equation}
{\partial\meanBB\over\partial t}=\etaT\nab^2\meanBB+\mu\nabla^2_\|\meanBB
+\epsilon\left(\nab_\perp^2\BB_\perp+\nab_\perp\nabla_\| B_\|\right),
\label{mu_aniso}
\end{equation}
which shows that positive values of $\mu$ correspond to an enhancement 
of turbulent diffusion along the preferred direction.
As \Eq{mu_aniso} reveals,
 $\eta_\|$ and $\eta_\perp$ do not characterize
the diffusion parallel and perpendicular to the 
preferred direction, as their symbols might suggest.

An anisotropy similar to that of \Eq{mu_aniso}
has been considered in connection
with the turbulent decay of sunspot magnetic fields \citep{RK00b},
where the mean magnetic field defines the preferred direction.
It has not yet been used in mean-field dynamo models, where, however,
anisotropies of the turbulent diffusivity due to
the simultaneous influence of rotation and stratification
have been taken into account
\citep{RB95,PK14} .

\section{The Model Setup}

We distinguish two basically different schemes of establishing the background flow:
by a prescribed forcing or by the convective instability.
In the first case, both
laminar and turbulent (artificially forced) flows will be considered.
With respect to the fluid, we generally think of an ideal gas
with state variables density $\rho$, pressure $p$ and temperature $T$, 
adopting, however, different effective
equations of state for the two schemes.

The continuity and induction equations are shared by both schemes and take the form
\begin{align}
{D\ln\rho\over D t} &=-\nab\cdot\UU, \label{eq:con}\\
{\partial \AAA\over\partial t} &=\UU \times \BB + \eta \nab^2 \AAA \, .  \label{eq:in}
\end{align}
Here $D/Dt = \partial/\partial t + \UU \cdot \bm\nab$ is the
advective time derivative and $\AAA$ is the magnetic vector potential.
The magnetic field includes the imposed field,
i.e., ${\BB} = {\BB}_{\rm ext} + \nab\times {\AAA}$,
and the microscopic diffusivity $\eta$ is constant.

\subsection{Forced flows}

In these models, we assume the fluid to be isothermal, which implies for
its equation of state $p=\cs^2 \rho$, with the constant sound speed $\cs$.
Hence we solve \Eqs{eq:con}{eq:in} together with the momentum equation,
\begin{align}
{D \UU\over D t} &=
-c_{\rm s}^2\nab\ln\rho
+ \rho^{-1} (\JJ \times \BB +\nab\!\cdot\!2\rho\nu\SSSS) + \ff.
\label{eq:vel}
\end{align}
Here $\nu={\rm const}$ is the kinematic viscosity,
and $\ff$ is a forcing function to be specified below.
The traceless rate of strain tensor $\SSSS$ is given by
\begin{equation}
{\sf S}_{ij} = \onehalf (U_{i,j}+U_{j,i}) - \onethird \delta_{ij} \DIV \bm{U},
\end{equation}
where the commas denote partial differentiation with respect to the
coordinate $j$.

The simulation domain for this model is periodic in all directions
with dimension 
$L_x \times L_x \times L_z$, that is, horizontally isotropic.
In the following we always use $L_x=L_y\equiv L$ and express lengths
in units of the inverse of the wavenumber $k_1=2\pi/L$.

\subsubsection{Roberts forcing}

First we use a laminar forcing to maintain 
one of the flows for which \cite{Rob72} had demonstrated dynamo action, namely his flow I.
It is incompressible, independent of  $z$,
and all 2nd-rank tensors obtained  from it by $xy$ averaging are symmetric about the $z$ axis \citep{Rae02a}.
The flow is defined by
\EQ
\uu_0= - \zz \times \nab \psi + k_{\rm f} \psi \zz \, ,
\label{UUrobflow}
\EN
where
\EQ
\psi = (u_0/k_0) \cos k_0x \, \cos k_0y \, , \quad k_{\rm f}=\sqrt{2}\,k_0\, ,
\label{UUrobflow2}
\EN
with constant $u_0$ and $k_0$.
Note that this flow is maximally helical, i.e. $\nab\times\uu_0=k_{\rm f}\uu_0$.
We define the forcing $\ff$ such that for $\BB=\bm{0}$, the 
flow \eqref{UUrobflow} with $\rho=\rho_0={\rm const}$
is an exact solution of \Eq{eq:vel}:
\begin{align}
\ff = \nu k_0^2 \uu_0+ \uu_0\cdot \nab \uu_0 \;\;
\left(= \nu k_0^2 \uu_0+ {\textstyle{1\over2}}\nab \uu_0^2\right).
\label{RobertsFlow}
\end{align}
We perform several simulations with different strengths of the
external magnetic field $\BBex$ 
with this flow.

\subsubsection{Forced turbulence}
\label{sec:forced}
Here we employ for $\ff$ a  random forcing function, namely
a linearly polarized wave with wavevector and phase being changed randomly between integration timesteps \citep{B01}.
The driven flow is non-helical and known to lack an $\alpha$ effect \citep{BRRK08}.
The averaged modulus of the wavevector is denoted
 by $k_{\rm f}$ and the ratio $k_{\rm f}/k_1$
is referred to as the scale separation ratio.
To achieve sufficiently large scale separation, we would need to keep 
$\kf/k_1$ large.
However, in this case $\Rm=\urms/\eta\kf$ becomes small.
Therefore we use $\kf/k_1 \ge 5$ as a compromise.

\subsection{Convection}

In this model the background flow is generated by convection
and consequently we employ
$p=(c_p-c_v)\rho T$ for the equation of state of the fluid,
where $c_p$ and $c_v$ are the specific heats at constant pressure and
volume, respectively.
Our model is similar to many earlier studies in the literature 
\citep[e.g.,][]{BJNRST96,OSB01,KKB08,KKB09a,KKB09b}.
Its computational domain is a rectangular box consisting
of three layers: the lower part ($-0.85 < z/d < 0$) is a convectively stable overshoot layer, 
the middle part ($0 \le z/d \le 1$) is convectively unstable, and the upper 
part ($1 < z/d < 1.15$) is an almost isothermal cooling layer.
The overshoot layer was made comparatively thick to guarantee that the overshooting is not
affected by the lower boundary. The box dimensions are $(L_x, L_y, L_z) = (5, 5, 2)d$, 
where $d$ is the depth of the unstable layer. Gravity is acting 
in the downward direction (i.e., along the negative $z$-direction).
By including rotation about the $z$ axis, we can consider the simulation box as a small portion 
of a star located at one of its poles. The mass conservation and induction equations  \eqref{eq:con}, \eqref{eq:in}
are now complemented by a modified momentum equation and 
an equation for the internal energy per unit mass \citep{BJNRST96}
\begin{align}
\hspace{-3mm} \frac{D\bm U}{Dt} &= -\frac{\bm{\nab}  p}{\rho} + {\bm g} - 2\bm{\varOmega} \times \bm{U} + \frac{1}{\rho} ( \JJ \times \BB + \bm{\nab} \cdot 2 \nu \rho \mbox{\boldmath ${\sf S}$}), \hspace{-3mm}\nonumber\\
\hspace{-3mm} \frac{De}{Dt} &= - \frac{p}{\rho}\bm{\nab}\!\cdot\UU + \frac{1}{\rho c_v} \bm{\nab}\! \cdot \! K(z) \bm{\nab}e + 2 \nu \mbox{\boldmath ${\sf S}$}^2 + \frac{\eta\mu_0}{\rho}\JJ^2 - \frac{e\!-\!e_0}{\tau(z)}. \hspace{-3mm}\label{equ:ene}
 \end{align}
Here, $\bm{g} = -g\hat{\bm{z}}$, with $g>0$, is the gravitational
acceleration, $\bm{\varOmega}=\varOmega_0(-\sin \theta,0,\cos \theta)$
is the rotation vector with $\theta$ its angle against the $z$ direction,
and $K$ is the heat conductivity
with a piecewise constant $z$ profile to be specified below.
The specific internal energy is related to the
temperature via $e=c_{v} T$.
In the energy equation \eqref{equ:ene}, the last term 
is of relaxation type and regulates the internal energy to settle on average close to $e_0={\rm const}$.
As there is permanent heat input from the lower boundary and from viscous heating, 
it effectively acts as a cooling. 
The relaxation rate $\tau(z)^{-1}$ has a value of $75 \sqrt{g/d}$ within the cooling layer
and drops smoothly to zero within the unstable layer
over a transition zone of width $0.025 d$. 

The vertical boundary conditions for the velocity are
chosen to be impenetrable and stress free, i.e.,
\begin{equation}
U_{x,z} = U_{y,z} = U_z = 0, 
\end{equation}
while for the magnetic field we use the vertical field boundary condition 
$B_x=B_y=0$. 
A steady influx of heat $F_0=-(K \partial_z e)(x,y,-0.85d)/c_v$
at the bottom of the box and a constant temperature,
i.e., constant internal energy, at its top is maintained,
where the latter is specified to be just equal to $e_0$ occurring in the relaxation term.
The $x$ and $y$ directions are periodic
for all fields. 

The input parameters are now determined in the following somewhat indirect way:
Instead of prescribing $K$, it is assumed that the hydrostatic reference solution 
coincides in the overshoot and unstable layers with a polytrope, the index $m$ of which is prescribed.
Here we choose  $m=3$ and $m=1$, respectively.
As for a polytrope $de/dz=-g/(m+1) (\gamma-1)$, $\gamma=c_p/c_v$, 
and at each $z$ we have 
$F_0= -(K/c_v) de/dz = {\rm const}$,
the heat conductivity is obtained as $K=c_v (m+1) (\gamma-1) F_0/g$, i.e., it is also piecewise linear
\citep[for a physical motivation, see][]{HTM86}. 
For simplicity it is assumed that in the cooling layer,
for which no polytrope exists,
$K$ has the same value as in the unstable one.
Within the ranges of the other control parameters covered by our
simulations, it is then guaranteed that the relaxation to the
quasi-isothermal state is dominated by the term $-e/\tau$.

The convection problem is governed by a set of dimensionless control
parameters comprising the Prandtl, Taylor, and Rayleigh numbers
\begin{align}
\Pra&=\frac{\nu}{\chi(z_{\rm m})},\ \ \Ta=\frac{4\varOmega^2 d^4}{\nu^2},\\ 
\Ra&=\frac{gd^4}{\nu \chi(z_{\rm m}) H_p(z_{\rm m})}\Delta\nabla(z_{\rm m}),
\end{align}
along with the dimensionless pressure scale height at the top
\begin{equation}
\xi_0 = (\gamma-1) e_0/gd.
\end{equation}
For the calculation of the Prandtl and Rayleigh numbers,
the values of the thermal diffusivity 
$\chi =K/\rho c_p$,
the superadiabatic gradient $\Delta\nabla=\dd(s/c_{\rm P})/\dd\ln p$ 
and the pressure scale height $H_p$
of the associated hydrostatic equilibrium solution
are taken from the middle of the convective layer at $z_{\rm m}=d/2$.
The density contrast within the unstable layer is
\EQ
{\rho(0)\over\rho(d)} = 1+ {gd\over2(\gamma-1)e_0} = 1 + {1\over2\xi_0}.
\label{eq:denscont}
\EN
Hence, the parameter $\xi_0$ controls the density
stratification in our domain. We use $\xi_0=3/25$ in all the
simulations which results in a (hydrostatic) density contrast of $31/6$;
$\gamma$ was fixed to $5/3$ throughout and
the different models have the same initial density at $z=z_{\rm m}$.

Equation \eqref{eq:denscont} assumes that 
$e=e_0$ at the top of the convective layer, which cannot be exactly true.
In the simulations this error is increased by the fact that the 
effect of the cooling reaches somewhat below $z/d=1$.
This leads to a higher density contrast ($\approx 8$) in the actual hydrostatic solution.

\subsection{Diagnostics}

As diagnostics we use the fluid and magnetic Reynolds numbers 
as well as the Coriolis number
\begin{equation}
\Rey= \frac{\urms}{\nu \kf}, \quad \Rm=\frac{\urms}{\eta \kf},
\end{equation}
where for the convection setup $\kf=2\pi/d$ is
an estimate of the wavenumber of the largest energy-carrying eddies.
$\urms =\langle \uu^2 \rangle^{1/2}$ is the rms value of the velocity with
$\langle\cdot\rangle$ denoting the average over 
the whole box or, for the convection setup, over the unstable layer
only, i.e.\ $0\le z/d \le 1$.
The values for $\Bex=0$, i.e., for the unquenched state, are marked by the subscript 0.
When we quote these values for a set of runs with different field strengths,
they apply to the case with weakest field, i.e., the unquenched state.
However for $\Rm$ we quote both the unquenched value, denoted by $\Rmz$,
and the quenched value for the run with the strongest field strength.

All simulations are performed using  the {\sc Pencil Code}%
\footnote{\url{http://pencil-code.googlecode.com}}, which uses
sixth-order finite differences in space and a third order
accurate explicit time stepping method.

\subsection{Test-field methods}
\label{sec:testfield}

The goal of the test-field method is to measure the turbulent transport coefficients 
completely from given flow fields
$\UU$, which can either be prescribed explicitly or produced by a numerical simulation, called the {\em main run}.
To accomplish this, the equation for the fluctuating fields 
\EQ
{\partial\aaaa^T\over\partial t}=\meanUU\times\bb^T+\uu\times\meanBB^T+
(\uu\times\bb^T)'+\eta\nab^2\aaaa^T 
\label{eq051}
\EN
is solved for a set of prescribed {\em test fields} $\meanBB^T\!\!$. Here $\bb^T=\nab\times\aaaa^T$
and the prime denotes the operation of extracting the fluctuation of a quantity.
Each $\aaaa^T$ results in a mean electromotive force
\EQ
\meanEMF{\,}^T=\overline{\uu \times \bb^T},
\label{eqtfemf}
\EN
and if the test fields are independent and their number is adjusted to that of the desired components in
$\alpha_{ij}$ and $\beta_{ijk}$,
the system \eqref{eqtfemf} can be inverted unambiguously.
For the truncated ansatz \eqref{meanemf1}, test fields which depend linearly on position are suitable.
However, when truncation is to be overcome, \Eq{meanemf1} can be considered as the Fourier space
representation of the most general $\meanEMF$--$\meanBB$ relationship.
Then $\alpha_{ij}$ and $\beta_{ijk}$ are functions of wavevector $\kk$ and angular frequency $\omega$
of the Fourier transform and it is natural to specify the test fields to be
harmonic in space \citep{BRS08} and time \citep{HB09}.
By varying their $\kk$ and $\omega$, arbitrarily close approximations
to the general $\meanEMF$--$\meanBB$ relationship
can be obtained \cite[see, e.g.,][]{Rae14}.

\subsubsection{Test-field method for horizontal ($z$-dependent) averages}

We will employ two different flavors of the test-field method.
For the first one we define mean quantities by averaging over all $x$ and $y$.
Then, necessarily, $\meanB_z={\rm const}$ and for homogeneous turbulence it is sufficient to consider
horizontal mean fields $\meanBB=\left(\meanB_x(z,t), \meanB_y(z,t), 0 \right)$ only.
When restricting ourselves to the limit of stationarity, our $\kk$ dependent test fields
have the following form:
\begin{eqnarray}
\meanv{B}^{1\rm c}&=&B_0(\cos kz,0,0), \;\;\; \meanv{B}^{2\rm c}=B_0(0,\cos kz,0),\quad\nonumber \\
\meanv{B}^{1\rm s}&=&B_0(\sin kz,0,0), \quad \meanv{B}^{2\rm s}=B_0(0,\sin kz,0),
\end{eqnarray}
where $k=k_z$ 
and in most of the simulations we use $k=k_1$. 
The $z$ component of  $\meanEMF$ does not influence $\meanBB$, thus
only its $x$ and $y$ components matter and we have
\begin{equation}
\meanemf_i=\alpha_{ij}\meanB_j - \eta_{ij}\meanJ_j,
\end{equation}
with $i,j = 1,2$ and
$\eta_{i1}= \beta_{i23}$ and $\eta_{i2}= -\beta_{i13}$. That is,
we can derive eight coefficients (four $\alpha$ and four $\eta$) using the above test fields.
Our main interest is to compute the diagonal components of $\alpha_{ij}$ and $\eta_{ij}$.
However, in some cases we also study the off-diagonal components.
Since the resulting turbulent transport coefficients depend
only on $z$ (in addition to $t$), we call this variant of
the test-field method \tfz.
It is implemented in the
{\sc Pencil Code} and discussed in detail by \citet{BRRS08}. 

It is convenient to discuss the results in terms of the quantities
\EQ
\begin{alignedat}{3}
\alpha&=&\onehalf(\alpha_{11}\!+\!\alpha_{22}),\;\;
\gamma&=\onehalf(\alpha_{21}\!-\!\alpha_{12}), \\
\etat&=& \onehalf(\eta_{11}\!+\!\eta_{22}), \;\; 
\delta &=\onehalf(\eta_{21}\!-\!\eta_{12}),
\end{alignedat}
\label{def:ae}
\EN
which cover an important subset of all the  eight coefficients.

\subsubsection{Test-field method for axisymmetric turbulence}
\label{sec:tfa}

Next, we turn to another variant of the test-field method that allows us
to calculate all nine coefficients in \Eq{axisymE} under the restriction
of axisymmetric turbulence.
It is then necessary to consider mean fields that depend on more than
one dimension,
as otherwise the gradient tensor $\nab\meanBB$ can be expressed
completely by the components of $\JJ$ and the coefficients $\kappa_\perp$,
$\kappa_\parallel$ and $\mu$ cannot be separated
from $\eta_\perp$, $\eta_\parallel$ and $\delta$.
Hence we now admit mean fields depending on all three spatial coordinates
and define the mean by {\em spectral filtering}.
We specify it such that only field constituents whose components 
$\sim \sum_j c_j(z) \exp \ii \overline{\kk}_j\cdot\xx_\perp$ 
contribute to the mean.
Here $\xx_\perp=(x,y)$ is  the position vector in horizontal planes
and the sum is over all two-dimensional wavevectors $\overline{\kk}_j$ of the form 
$(\pm \overline{k}_{x},\pm \overline{k}_{y})$ with fixed 
$\overline{k}_x, \overline{k}_y > 0$.
So averaging means here to perform the operation
\begin{equation}
     \hspace{1mm} \overline{f}(x,y,z) =  \frac{1}{A} \sum_j \!
     \!  \int_A  f(x',y',z)  {\rm e}^{\ii \overline{\kk}_{j}\cdot(\xx_\perp-\xx_\perp')} dx' dy' ,     \hspace{-4mm}
\end{equation}
where $A$ is the horizontal cross-section of the box.
We call this variant of the test-field method for axisymmetric turbulence \tfa\ and
refer for further details to \cite{BRK12}.
In our case, the preferred direction is given by that of
the externally imposed magnetic field. 

As we will apply this method only with 
horizontally isotropic 
periodic
boxes with $L_x=L_y=L$, we may choose $\overline{k}_{x}=\overline{k}_{y}=k_1=2\pi/L$.
In general, it makes not much sense to choose $\overline{k}_{x}$ or $\overline{k}_{y}$  different from these smallest possible values for the corresponding extents of a given box.
Otherwise, possible field constituents with smaller wavenumbers would
be counted to the `fluctuations' which is hardly desirable. 
Even for our choice $\overline{k}_{x},\overline{k}_{y}=k_1$
this could be a problem, namely with respect to 
constituents with
horizontal wavenumber $k_x$ or $k_y$ equal to zero, 
so  their occurrence should be avoided. 
As we apply \tfa\ only to homogeneous turbulence (fully periodic boxes), this is granted.

Spectral filtering, being clearly useful for comparisons with observations,
is known to violate in general the Reynolds rule $\overline{\overline{F}g}=0$.
However, if in the $\kk$ spectrum of the quantity $G=\overline{G}+g$
there are  ``gaps'' at $k_x, k_y  = 2 k_1$
(for our choice) with vanishing spectral amplitudes, this rule is granted%
\footnote{To let \Eq{eq051} hold, we need also $\meanUU=\bm 0$,
otherwise a term $(\meanUU\times\meanBB)'$ would show up.}.
Such gaps, albeit only in the form of amplitude depressions,
can emerge in the saturated stage of a turbulent dynamo;
see \cite{B01} for examples, where this phenomenon was
characterized as ``self-cleaning''.
In the kinematic stage, on the other hand, gaps cannot be expected
and it remains unclear to what extent a mean-field approach,
based on spectral filtering, can then be useful.

In this method, we use four test fields defined by
\EQ
\hspace*{-2mm}
\begin{aligned}
\meanBB^{\rm 1c} &\!= (B_0 \cx \cy \cz, 0, 0), \;\meanBB^{\rm 1s}  = (B_0 \cx \cy \sz, 0, 0), \\
\meanBB^{\rm 2c} &\!= (0, 0, B_0 \cx \cy \cz), \; \meanBB^{\rm 2s} = (0, 0, B_0 \cx \cy \sz),
\end{aligned}\hspace*{-3mm} \label{eq:axtf}
\EN
where $B_0$ is a constant and we have used the abbreviations
\EQ
\begin{alignedat}{2}
\cx &=&\, \cos \overline{k}_x x \, , \quad \cy &= \cos \overline{k}_y y\, , \\ 
\cz &=& \cos k_z z \, , \quad \sz &= \sin k_z z.
\end{alignedat}
\EN
Note the different roles of the wavenumbers: while
$\overline k_x$ and $\overline k_y$ are {\em defining} the mean,
by $k_z$ a specific mean field out of the infinitude of possible ones
is selected.\footnote{Horizontal averaging with \tfz\ is equivalent to
spectral filtering with \tfa\ using $\overline{k}_x=\overline{k}_y=0$.
In that special case, all Reynolds rules are obeyed.
In \cite{BRK12}, $\cx$ and $\cy$ were replaced by
$\sin \overline{k}_x x$ and $\sin \overline{k}_y y$, respectively.
This is equivalent to the former, except that then \tfz\ cannot be
recovered for $k_x=k_y=0$.}
Other than what could be expected, three test fields are in general not
sufficient to calculate the wanted nine coefficients,   
as the linear system from which they are obtained suffers from a rank defect.
For {\em homogeneous} turbulence, however, exploiting the orthogonality
of the harmonic functions, even only two test fields were sufficient.

\subsubsection{Computing transport coefficients via resetting}

At large $\Rm$, we often find the solutions $\aaT$ of the test problems \eqref{eq051}
to grow rapidly due to the occurrence of unstable eigenmodes of their homogeneous parts.
Therefore, similar to earlier studies \citep{SBS08, MKTB09, KKB09a,HDSKB09},
we reset $\aaT$ to zero after a certain time interval to prevent the unstable eigenmodes from dominating
and thus contaminating the coefficients.
If the growth rates are not too high, after an initial transient phase `plateaus' can be identified 
in the time series of the coefficients, during which 
they are essentially determined by the {\it bounded} solutions of 
the (inhomogeneous) problems  \eqref{eq051}.
Even for monotonically growing $\aaT$, sufficiently long plateaus
occur as the averaging in
 the determination of $\meanEMF{\,}^T$ (\Eq{eqtfemf}) is capable of eliminating the unstable eigenmodes. 
Typically we use data from ten such plateaus to
compute the temporal averages of the transport coefficients
and ensure by spot checks that the results do not depend on the length of the resetting interval.

\section{Results}
\label{sec:results}

\subsection{Roberts flow}
We describe here results for Roberts~I flow forcing for two different
parameter combinations (Sets~RF1 and RF2 in Table~\ref{tab:runs}).
For RF1 we choose $u_0 = 0.01 \cs$, and $\eta= \nu =0.008 \cs/k_1$, whereas for RF2 
$u_0 = \cs$ and $\eta = \nu =\cs/k_1$, using $k_0=k_1$ for both.
With a vertical field, $\BBex = \Bex \zu$, we have
$\nab\times(\uu_0\times\BBex)
=\BBex\cdot\nab \uu_0
=\bm 0$.
Hence there is no tangling of the field and consequently no
effect of $\BBex$ on the flow, that is, no quenching.
(Should, however, the flow have undergone a bifurcation and thus deviate
from \eqref{UUrobflow}, this needs no longer be true.
However, for the fluid Reynolds numbers considered in this paper
we have not noticed any bifurcations.)
Therefore we choose a horizontal field $\BBex = \Bex \xu$.   
We apply the \tfz\ procedure with test-field wavenumber $k=k_1$
and normalize the resulting $\alpha_{ij}$ and $\eta_{ij}$ by 
the corresponding SOCA results in the limit of $k\rightarrow 0$
\EQ
\alpha_0 = - u_0^2 / (2 \eta \kf) \, , \quad  
\eta_{\rm t 0} =  u_0^2 / (2 \eta \kf^2) \,,
\label{eq52}
\EN
see \cite{Retal14}.

\begin{figure}[t]
\centering
\hspace*{-0.5mm}\includegraphics[width=1.06\columnwidth]{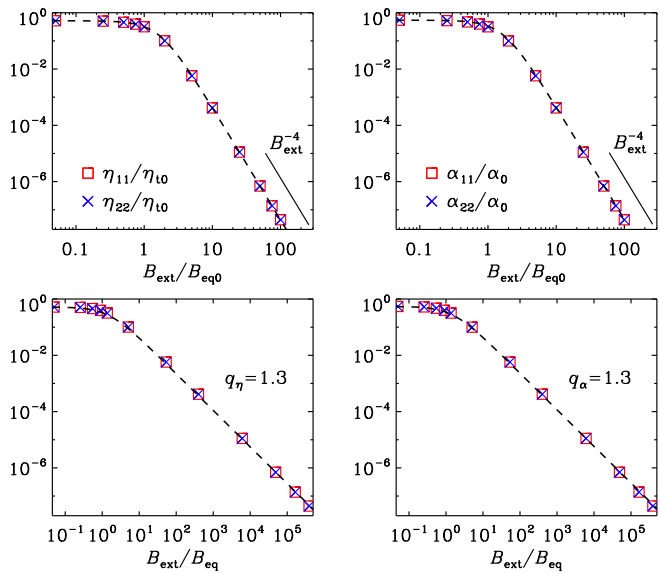}
\caption{Results of Set~RF1, $\Rmz=0.88$:
Variations of $\eta_{11}$, $\eta_{22}$ (left) and $\alpha_{11}$,
$\alpha_{22}$ (right) with the external magnetic field in the $x$ direction.
Top: dependences on $\Bex/\Beqh$ with fit \eqref{scalling2}.
Bottom: dependences on $\Bex/\Beq$ with fit \eqref{scalling}.
Fits dashed.}
\label{fig:roberts}
\end{figure}

We compute $\Beqh =\urmsh \langle \rho \rangle^{1/2}$ from a simulation
without external magnetic field (the ``hydro run'', subscript ``0'')
or by virtue of \Eq{RobertsFlow} directly from the forcing amplitude.
In \Fig{fig:roberts} we show the diagonal components of
$\alpha_{ij}$ and $\eta_{ij}$ for Set~RF1
as functions of $\Bex/\Beqh$ and also of $\Bex/\Beq$,
where $\Beq$ is derived from the actual $\urms$ and hence dependent on $\Bex$.
The off-diagonal components are zero to high accuracy  
while $\alpha_{11}$ and $\alpha_{22}$ as well as $\eta_{11}$
and $\eta_{22}$ are very close to each other (to four digits).
This apparent isotropy of the {\em quenched} flow is somewhat surprising as the imposed field
is in general capable of introducing a new preferred direction. So let us consider the second-order change in the flow, $\uu^{(2)}$,
for small $\Bex$. From \eqref{eq:in} we get for the first-order magnetic fluctuation in the stationary case
and with $\Rm\ll 1$
\EQ
    \hspace{-0mm}   \bb^{(1)} = \frac{\Bex}{\eta k_0} (v_0 sx\, sy, v_0 cx\, cy, -w_0 sx\, cy), 
\EN
and from this the solenoidal part of the second order Lorentz force
\EQ
    \BBex\!\cdot\!\nab \bb^{(1)} = 
   \frac{\Bex^2}{\eta}(v_0 cx \, sy, -v_0 sx\, cy, -w_0 cx\, cy) \sim \uu_0  .
\EN
A quadratic contribution  from $\bb^{(1)} $ is not present as the Beltrami property 
$\nab\times \bb^{(1)}  \sim \bb^{(1)} $ holds. 
That is, if the Reynolds numbers (here $\le 0.88$) as well as the modification of the pressure 
(and thus density) by the magnetic contribution $\bb^{(1)}\cdot \BBex $ is small 
(i.e., if the corresponding plasma beta is large), the 
flow {\em geometry} is not changed by its second-order correction.
This implies that our argument continues to hold up to arbitrary orders in $\Bex$, 
if only the $\uu\cdot\nab \uu$
and $\nab\times(\uu\times\bb)'$
terms can be neglected and the pressure modification is small at any order.
So for our values of $\Rm$ and
$\Rey$ the quenched flow differs from the original one mainly in 
amplitude and preserves essentially its horizontal isotropy. 

The condition for $\Rey$ can be relaxed when rewriting 
the advective terms in the second-order momentum equation
in the form $\left(\nab \times \uu^{(2)}\right) \times \uu_0 + (\nab \times \uu_0) \times \uu^{(2)} + \nab\left( \uu^{(2)} \cdot \uu_0 \right)$.
A solution $\uu^{(2)}  \sim \uu_0$ can already exist (approximately), if
the sum of magnetic and dynamical pressure,
 $\bb^{(1)} \cdot \BBex +  \rho_0 \uu^{(2)}  \cdot \uu_0$,
is negligible compared to 
$p_0 = \cs^2 \rho_0$,
more precisely and less restricting, if the non-constant part  of this sum is negligible.
At higher orders there is an increasing number of contributions to be taken into account.

Therefore, following the definition \eqref{def:ae}, we can define
$\alpha \approx \alpha_{11} \approx \alpha_{22}$ and 
$\eta_{\rm t} \approx \eta_{11} \approx \eta_{22}$.
The transport coefficients start to be quenched when $\Bex$ exceeds $\Beqh$ or $\Beq$ and seem 
to follow a power law for strong fields. To compare with 
earlier works, it is useful to consider the dependences on $\Bex/\Beqh$.
Calculating the quenched coefficients under SOCA by a power series expansion with respect to $\Bex$,
only the even powers occur.
Accordingly,
we find that our data fit remarkably well with
\EQ
   \sigma=\frac{\sigma_0} {1 + p_{\sigma1} (\Bex/\Beqh)^2 + p_{\sigma2} (\Bex/\Beqh)^4} \,,
\label{scalling2}
\EN
where $\sigma$ stands for $\alpha$ or $\etat$ and $p_{\sigma1}=0.51$ and $p_{\sigma2}=0.12$;
see upper panels of \Fig{fig:roberts}.
Therefore our results are consistent with those of \cite{SBS07} and \cite{RB10} who found 
asymptotically the power $4$
for steady forcing.\footnote{In \citet{SBS07} a leading power of $3$
is quoted, but the data in their Figure~2 are actually closer to a 
power of $4$ as was already pointed out in Sect.~4.2.1 of \cite{RB10}.}

Alternatively, we may consider the dependences on $\Bex/\Beq$ 
which are weaker, because the actual $\Beq$ is itself quenched.
We find as an adequate model
\EQ
   \sigma=\frac{\sigma_0} {1 + p_\sigma (\Bex/\Beq)^{q_\sigma}} \quad\mbox{(for $\sigma=\alpha$ or $\etat$)}
\label{scalling}
\EN
with $q_\alpha \approx q_\eta \approx 1.3$ and $p_\alpha \approx p_\eta \approx 0.59$; see lower panels
of \Fig{fig:roberts}.
From now onward we shall consider the dependences on $\Bex/\Beq$
and stick to the fitting formula~(\ref{scalling}). 
We have performed another set of simulations with different parameters
(RF2 in Table~\ref{tab:runs}) 
and also at different wavenumbers of the test-fields.
In all the cases we get the same quenching behavior.

The obtained isotropy of the quenched coefficients seems to be in conflict with the results of \cite{RB10},
who detected strong anisotropy in $\alpha_{ij}$ for Roberts~I forcing.
However, the analytic consideration above makes clear, that this was a consequence of their use
of a simplified momentum equation missing the pressure term.
Thus, the ingredient just necessary to allow the flow keeping its geometry while being influenced by the imposed field, was missing. 
One may speculate though, that for more compressive flows the anisotropy may become larger.

\begin{deluxetable*}{@{\hspace{1mm}}ll@{\hspace{0mm}}c@{\hspace{1.5mm}}rc@{\hspace{1mm}}r@{\hspace{0mm}}c@{\hspace{1mm}}c@{\hspace{0.5mm}}r@{\hspace{0mm}}c@{\hspace{1.5mm}}cccccccc}
\tabletypesize{\scriptsize}
\tablecaption{Summary of the runs.
}\tablecomments{
Data given for the 
stationary (Sets~RF1 and RF2) or statistically saturated state, 
respectively. $q_{\alpha}$, $q_{\gamma}$, $q_{\eta}$,
and $q_{\delta}$ are the
quenching exponents for $\alpha$, $\gamma$, $\etat$, and $\delta$, respectively, 
according to \Eq{scalling}. For 
RF1: $u_0=0.01\cs,\eta=0.008 \cs/k_1$,
RF2: $u_0=1.0\cs,\eta=\cs/k_1$,
CR0: $\Ra,\Pr=3 \times 10^5, 3.95$,
CR1: $\Ta,\Ra,\Pr=5.6 \times10^3, 3 \times 10^5, 3.95$, 
CR2: $\Ta,\Ra,\Pr=6.4\times10^5, 3 \times 10^5, 4.94$,
CR3: $\Ta,\Ra,\Pr=1.0\times10^4, 4 \times 10^5, 2.93$,
CR4: $\Ta,\Ra,\Pr=2.6\times10^6, 5 \times 10^5, 2.44$,
CR5: $\Ta,\Ra,\Pr=1.6\times10^7, 1 \times 10^6, 0.97$.
Resolutions used are
RF1: $96^3$,
RF2: $144^3$,
TBx, TBz: $256^3$,
AT1: $128^3$,
AT2: $360^3$,
AT3: $72^3$ to $672^3$
CR0-CR7: $128^3$.
$\Rm^{\rm min}$ -- minimal, i.e., maximally quenched $\Rm$ within a Set.
}
\tablewidth{0pt} \tablehead{
Set  & Description                     &${\hat \BB}_{\rm ext}$&TFM&$\Rmz\;$ & $\Rm^{\rm min}$&& $\:\Pm$ & $\Beqh$  && $p_{\alpha}$ & $p_{\gamma}$ & $p_{\eta}$ & $p_{\delta}$ & $q_{\alpha}$ & $q_{\gamma}$ & $q_{\eta}$ & $q_{\delta}$
}\startdata
RF1  & Forced Roberts flow                         &$\xu$ &  \tfz  & 0.88&0.0002&& 1.0 & 0.010  && 0.59& -  & 0.59& - & 1.3  & -  & 1.3  &  - \\      
RF2  & Forced Roberts flow                         &$\xu$ &  \tfz  & 0.707&0.10  && 1.0 & 1.000  && 0.3  & -  & 0.4  & -  & 1.3  & -  & 1.3  &  - \\[0.75mm]
\hline\\[-1.5mm]
TBx  & Forced turbulence ($\kf=5 k_1$)    &$\xu$ &  \tfz  & 0.87 &0.71 && 1.0 & 0.045   &&  -    & -  & 0.38 &  -  &  -  & -  & 1.1  &  -      \\
TBz  & Forced turbulence ($\kf=5 k_1$)    &$\zu$ &  \tfz  & 0.87 &0.71 && 1.0 & 0.045   &&  -    & -  & 0.21 &  -  &  -  & -  & 1.1  &  -      \\[0.75mm]
\hline\\[-1.5mm]
AT1  & Forced turbulence ($\kf=27 k_1$) &$\zu $&  \tfa  & 2.23&1.67  && 1.0 & 0.060   &&  -    & -  & 0.66 &  -  &  -  & -  & 1.2  &  - \\
AT2  & Forced turbulence ($\kf=27k_1$)  &$\zu $&  \tfa  & 18.2&15.8  && 1.0 & 0.100   &&  -    & -  & 2.50 &  -  &  -  & -  & 1.0  &  -  \\[0.75mm]
\hline\\[-1.5mm]
AT3  & Forced turbulence ($\kf=27 k_1$),              &$\zu $&  \tfa &&   0.08 && 1.0 & 0.022 &&     -  &  -  & -  &   -  &  -  &  -  & - & - \\ 
         &\phantom{Fo}$\Bex/\Beq\approx 4.3$ fixed&          &          && -- 537&\footnote[$\dagger$]{\scriptsize Not the minimum, but the range of values of the individual runs.}  & 
                                                                                                                                               &-- 0.116&\footnote[$\ddagger$]{\scriptsize $\Beq$ instead of $\Beqh$.}&  \\[0.75mm]
\hline\\[-1.5mm]
CR0  & Non-rotating convection           &$\zu$&   \tfz &  3.91&0.5   && 0.8 & 0.087 &&  -   & 0.34 & 0.34 &  -   &  -     & 1.2   & 1.2   &  -      \\
CR1  & Rotating convection               &$\xu$&   \tfz &  3.85&0.7   && 0.8 & 0.087 && 0.11 & 0.12 & 0.20 & 0.02 & 1.8  & 1.4   & 1.3   & 2.0   \\
CR2  & Rotating convection               &$\xu$&   \tfz & 11.7 &0.2   && 5.0 & 0.054 && 0.14 & 0.11 & 0.10 & 0.06 & 1.8  & 1.3   & 1.3   & 1.8   \\
CR3  & Rotating convection               &$\xu$&   \tfz & 20.2 &6.0   && 3.0 & 0.090 && 0.25 & 0.24 & 0.65 & 0.06 & 2.0  & 1.3   & 1.3   & 2.0   \\
CR4  & Rotating convection               &$\xu$&   \tfz & 29.1 &3.3   && 5.0 & 0.065 && 0.17 & 0.15 & 0.16 & 0.07 & 2.0  & 1.3   &  1.3  & 1.8   \\
CR5  & Rotating convection               &$\xu$&   \tfz & 89.5 &19.2  && 5.0 & 0.082 && 0.24 & 0.23 & 0.37 & 0.12  & 1.8  & 1.24 & 1.26 & 1.7  \\
CR1Bz& As CR1                            &$\zu$&  \tfz  & 3.85 &0.13  && 0.8 & 0.088 && 0.65 & 0.12 & 0.41 & 1.0   & 1.3  & 1.3   & 1.3    & 1.8  \\
CR3Bz& As CR3                            &$\zu$&   \tfz & 28.6 &0.08  && 5.0 & 0.065 && 0.59 & 0.21 & 0.41 & 0.25  & 1.3  & 1.4   & 1.3    & 2.0  \\
CR6  & As CR1, uniform test fields       &$\xu$&   \tfz & 3.85  &0.70 && 0.8 & 0.087 && 0.16 & 0.11 &  -   &  -    & 2.0  & 1.4   &  -        &  -     \\
CR7  & As CR2, uniform test fields       &$\xu$&   \tfz & 29.1  &3.2  && 5.0 & 0.065 && 0.16 & 0.16 &  -   &  -    & 2.0  & 1.3   &  -        &  -  
\enddata
\label{tab:runs}
\end{deluxetable*}

\subsection{Stochastically forced turbulence}

Previous work using stochastically forced turbulence has mainly focussed
on $\alpha$ using the imposed-field method \citep{BNPST90,CH96,HDSKB09}.
An exception is the work of \cite{BRRS08}, where $\alpha$ and $\etat$
have been determined simultaneously using \tfz\ for super-equipartition
magnetic fields resulting from saturated dynamo simulations in a triply
periodic domain.
Here we employ the non-helical stochastic forcing described in
\Sec{sec:forced} with a strength adjusted such that the flow remains
subsonic (Mach number $\approx0.1$).
We have performed several simulations with different values of $\Rmz$ 
and with different orientations of
$\BBex$.
Both \tfz\ and \tfa\ are applied to measure
the turbulent transport coefficients.
For the latter we considered the requirement of  gaps in the spectra of the fields (see \Sec{sec:tfa}) by choosing a high forcing wavenumber, $\kf=27 k_1$.

Due to the imperfectness of isotropy and homogeneity caused by finite
scale separation of the forcing, the coefficients show fluctuations both
in space and time.
We usually remove them by averaging over the whole box and sufficiently long times.
An exception are the coefficients $\alpha_\perp$ and $\alpha_\|$ that vanish on average
owing to the lack of helicity, but whose
fluctuations are still of interest; see \Sec{sec:incoh}.
As expected, and in agreement with earlier work \citep{BRK12},
$\gamma$ and $\delta$ also vanish on average and are not shown here.

The time spans for temporal averaging should ideally be so long that the averages become stationary. 
How close we came to this is
in several cases indicated by error bars showing the largest deviation of the average over any
one third of the time series from the overall average. 

It is convenient to normalize the results using the unquenched, hence
isotropic expression $\etat$ as obtained in SOCA
in the high conductivity limit, i.e.\
\begin{eqnarray}
\etatz=\onethird \urms \kef^{-1}.
\label{alpeta0}
\end{eqnarray}
When we determine the fluctuations of $\alpha$, we use $\alpha_0=\urms/3$
for normalization which would be the 
expected value in fully helically forced isotropic turbulence.
First we present the transport coefficients measured using
\tfz, but restrict ourselves to $\eta_{11}$ and $\eta_{22}$.

\subsubsection{\tfz: horizontal and vertical fields}

\begin{figure}[t]
\vspace{5mm}
\centering
\includegraphics[width=0.50\textwidth]{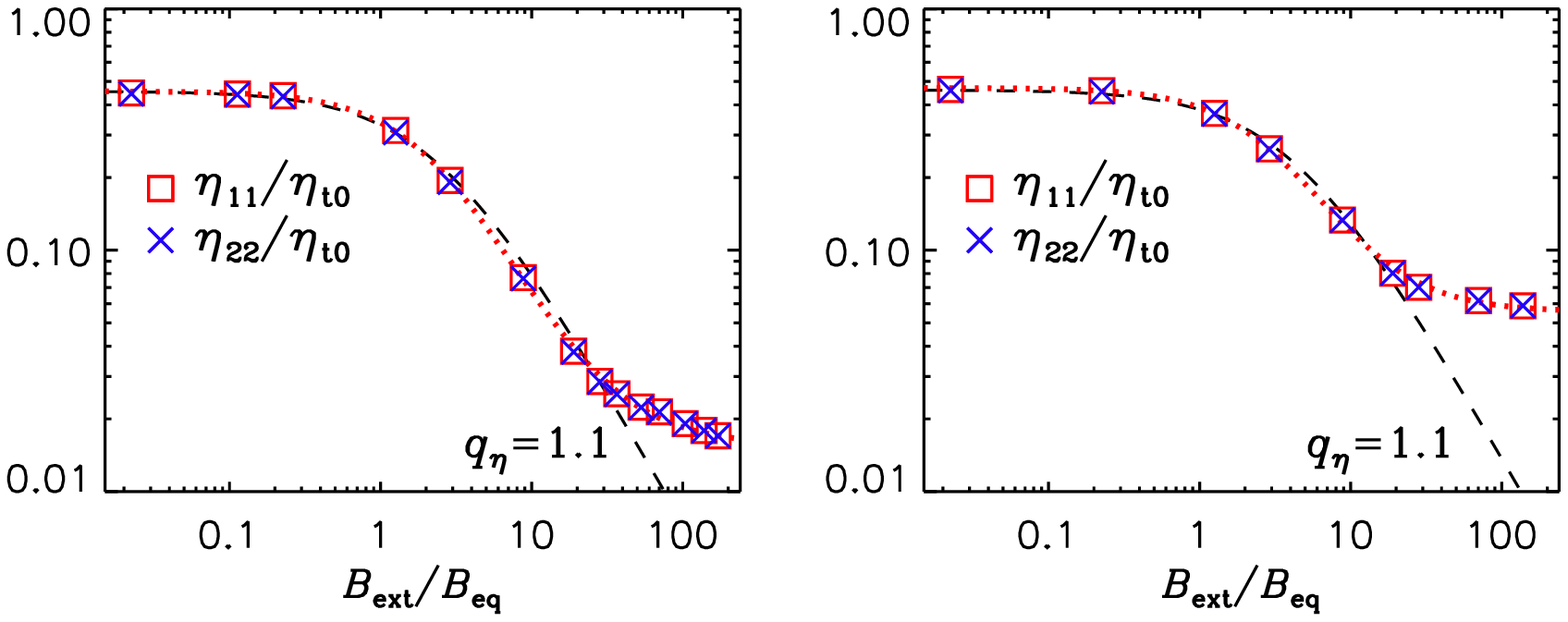}
\caption{Dependence of $\eta_{11}$ and $\eta_{22}$ on the
imposed field for forced turbulence; $\Rmz = 0.87$.
Left: Set~TBx with $\BBex \parallel \xu$,
right: Set~TBz with $\BBex \parallel \zu$.
Dashed lines: fits from \Eq{scalling} with exponents $q_\eta$;
dotted lines: fits from \Eq{eq:mrfunc}.}
\label{fig:etaquench1}
\end{figure}

Figure~\ref{fig:etaquench1} shows the results for both horizontal 
and vertical external fields, $\BBex=\Bex \xu$ (Set~TBx)
and $\BBex=\Bex \zu$ (Set~TBz); see Table~1.
For these runs we have adopted $\kf/k_1 = 5 $ and
$\eta=\nu=0.01\,\cs/k_1$ which yields $\Rmz = 0.87$.
Note that in both cases $\eta_{11}$ is almost identical to $\eta_{22}$,
which is natural for the vertical field, but unexpected for the horizontal one,
because $\eta_{ij}$, being an axisymmetric rank-2 tensor whose preferred direction
is given by $\hatmeanBB \parallel \BBex$, must have the general form
$\eta_{ij}=\eta_0\delta_{ij}+\eta_1\hatmeanB_i\hatmeanB_j$
with $\meanB$-dependent coefficients $\eta_0$ and $\eta_1$.
This has indeed been confirmed previously for 
a dynamo-generated $\meanBB$ of Beltrami type \citep{BRRS08}.
For horizontal $\BBex$ we have thus $\eta_{11}= \eta_0+ \eta_1$, but $\eta_{22}=\eta_0$.
The reason for the apparent vanishing of $\eta_1$ is currently unclear,
but might be connected with the fact that here the field is a uniform one.

Indeed, considering a forcing, simplified such that only a single transverse (frozen) wave is supported
instead of switching rapidly between waves with random wavevector and phase, one finds that 
a uniform imposed field of arbitrary strength does not change the geometry of that wave, but merely its amplitude,
see \App{app1}. Hence,
for a statistical ensemble, generated by random choices of wave and polarization vectors, $\eta_{ij}$ from averaging over this ensemble must remain
isotropic, that is,  
$\eta_1$ needs to vanish.
The only condition for that to hold is the negligibility of the pressure variations caused by the imposed field, 
compared to the pressure in the field-free case. This finding looks similar to that obtained for the Roberts forcing case,
although the mathematical reason is  here the transversality of the wave flow and not its Beltrami property.
Returning to the actually used 
delta-correlated 
random-wave forcing, one would conclude, that approximate isotropy could occur
as long as the waves are damped quickly enough for letting their mutual interaction be subdominant.
Of course, 
if at all,
this can only happen for small $\Rey$ and $\Rm$ as those in Sets~TBx and TBz ( $\Rey_0=\Rmz=0.87$).
With increasing Reynolds numbers, anisotropy should gradually emerge, and indeed, for $\Rey_0=\Rmz\approx 14$
we find $\eta_{11}$ being by $\approx 9\%$ bigger than $\eta_{22}$ when the imposed field is as weak as $\Bex/\Beq=0.66$.

Unlike the Roberts flow case, the behavior for $\Bex > \Beq$ cannot be described by
a single asymptotic power law.
Instead we observe a possible transition from one power law 
to another one with lower power at $\Bex \gtrsim 20 \Beq$.
Accordingly, the fitting formula \eqref{scalling}, with quenching
exponents $q_\eta=1.1$ for both cases matches well only up to this value.
Nevertheless in \Fig{fig:etaquench1} we see that the quenching is not 
exactly the same for the two field directions, namely slightly weaker for the 
vertical field as we find $p_\eta=0.21$ for the latter, but $p_\eta=0.38$ for the horizontal field.

A satisfactory overall fit can be obtained by employing an ansatz of the form
\EQ
\eta_{11,22}(\Bex) = \eta_{11,22}(0)
\frac{1+p_{\rm n} (\Bex/\Beq)^{q}}{1+p_{\rm d} (\Bex/\Beq)^{q}}   \label{eq:mrfunc}
\EN
with $q=1.36$ and $1.31$ for horizontal and vertical field, respectively; see the red dotted lines in \Fig{fig:etaquench1}.
This can be taken as an indication of asymptotic independence of $\eta_{11,22}$ on $\Bex$,
which makes sense as the turbulence should asymptotically become two-dimensional with $\BBex\cdot\nab \uu = \bm 0$.  
Note that we do not observe this in the Roberts forcing case
because there, as demonstrated above, the flow has no freedom to adjust to this condition,
at least for not too high $\Rm$.

If we normalize $\Bex$ by $\Beqh$, the scaling for the quenching
changes and the exponent $q_\eta$ becomes $1.5$ and $1.4$
for horizontal and vertical external fields, respectively.
These values are higher than the result of \citet{KPR94} and \citet{RK01}, who found 
unity.

When comparing the two panels of \Fig{fig:etaquench1} one might ask why the quenching characteristics of $\eta_{22}$ for horizontal and  vertical $\BBex$ are
not identical although this coefficient is in both cases
correlating components of $\meanEMF$ and $\meanJJ$ perpendicular to the preferred direction.
This apparent ambiguity can be resolved with a view to \Eq{mu_aniso}: Provided that $\epsilon\approx 0$
(which will be demonstrated in the next section),
we have for vertical external field $\nab = \nabla_\| \eee_z$,
hence $\etat$ and $\mu$ sum up, while for horizontal $\BBex$
of course $\nabla_\|=0$, so $\eta_{11}(=\eta_{22})$
should differ in the two cases roughly by $\mu$.
That is, the anisotropy of the turbulence does manifest in the diffusive
behavior, but not by causing an anisotropic $\eta_{ij}$.
 
\begin{figure}[t]
\centering
\hspace*{-1mm}\includegraphics[width=0.5\textwidth]{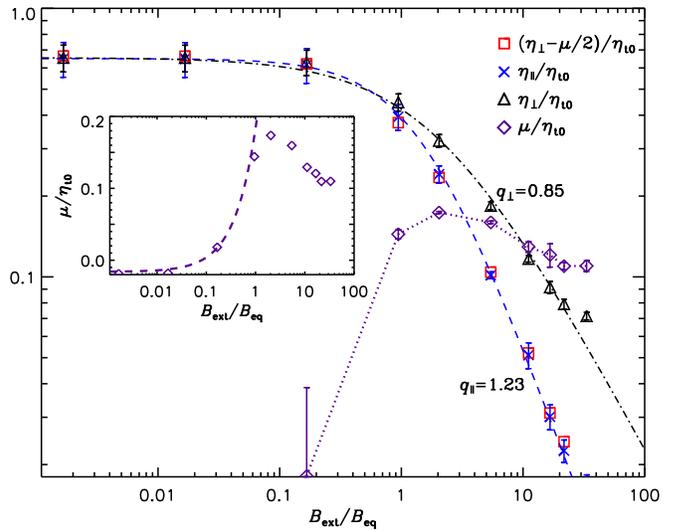}
\caption{Results of Set~AT1, $\Rmz = 2.23$.
Variation of $\eta_\parallel$ (crosses), $\eta_\perp$ (triangles),
$\mu$ (diamonds, dotted) and $\eta_\perp - \mu/2$ (squares), all normalized by $\etatz$, with the imposed field 
$\Bex/\Beq$; 
dashed and dash-dotted: fits 
to $\eta_\parallel$ and $\eta_\perp$, respectively, from 
\Eq{scalling}. Dashed line in inset: linear fit with slope $0.2$.}
\label{fig:AT1}
\end{figure}

\subsubsection{\tfa: determining anisotropy}

To measure the anisotropy of turbulent diffusion,
we have applied \tfa\ for axisymmetric turbulence whose
preferred direction is defined by the imposed field.
Hence, we consider the case $\BBex =\Bex\zu$.
We measure all the relevant transport coefficients described in \Eq{axisymE}. 
Here we only show $\eta_\perp$, $\eta_\parallel$, and $\mu$.
It turns out that $\kappa_\perp$ and $\kappa_\|$ are negative
(around $-0.01$ in units of $\etatz$) for our largest field strengths,
but zero within error bars for weaker fields and hence not shown.
All other coefficients are at least about ten times smaller and
fluctuating about zero; see \Sec{sec:incoh} for 
some discussion about those fluctuations.
We denote this set of simulations by AT1
and show its results in \Fig{fig:AT1}; see also Table~1.
It turns out that 
$\eta_\perp$ is less strongly quenched than $\eta_\parallel$.
According to the fitting formula \eqref{scalling}, 
we have $q_\parallel = 1.2$, but $q_\perp=0.85$.
The coefficient $\mu$ is increasing with $\Bex$
until a maximum at  $\Bex/\Beq\approx 2$. 
Interestingly, we have $\eta_\parallel\approx\eta_\perp-\mu/2$; see  
red squares in \Fig{fig:AT1}. 
If we apply this finding in \Eq{mu_aniso} we see that because of $\epsilon\approx 0$ 
the mean-field induction equation takes the simple form $\partial_t \meanBB = \big((\eta + \eta_\perp - \mu/2) \nab^2 + \mu \nabla_\|^2\big)\meanBB$.
We may redefine the preferred direction to coincide now with $\xxx$ and assume at the same time, that all mean quantities depend solely on $z$, hence $\nabla_\| = 0$.
In this way we can make contact with the results of 
\tfz\ for horizontal fields arriving at $\eta_{11}=\eta_{22}=\eta_\perp -\mu/2$.
So the somewhat surprising {\em isotropy} of $\eta_{ij}$ obtained with \tfz\ is confirmed 
with \tfa\ in spite of $\eta_\perp \ne \eta_\|$.

\begin{figure}[t]
\centering
\hspace*{-1mm}\includegraphics[width=0.51\textwidth]{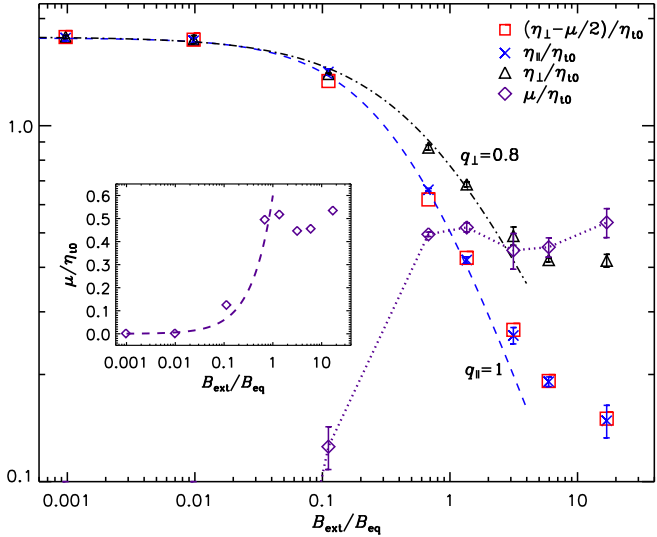}
\caption{As \Fig{fig:AT1}, but for Set~AT2, $\Rmz=18.2$ and the linear fit
(dashed line in inset) here has slope $0.6$.}
\label{fig:AT2}
\end{figure}

\begin{figure}[t]
\centering
\includegraphics[width=.9\columnwidth]{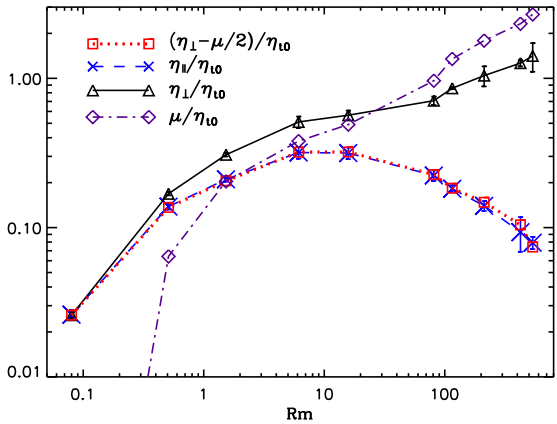}
\caption{$\Rm$ dependence of the turbulent diffusivity in
axisymmetric turbulence with a fixed $\Bex/\Beq \approx 4.3$.
Squares:
$\eta_\perp-\mu/2$;
crosses:
$\eta_\parallel$; 
triangles
$\eta_\perp$;
diamonds:
$\mu$, all normalized by $\etatz$.}
\label{fig:rm}
\end{figure}

\begin{figure}[t]
\centering
\includegraphics[width=0.5\textwidth]{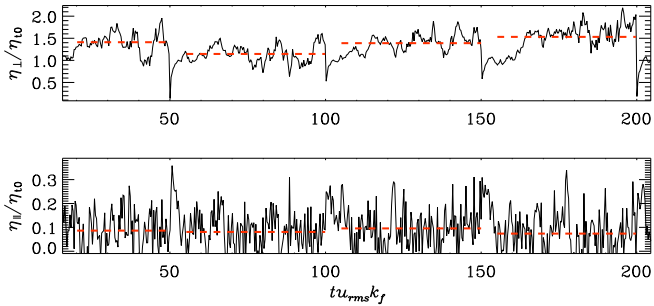}
\caption{A small portion of the time series of $z$-averaged $\eta_\perp$
and $\eta_\parallel$ for the highest $\Rm=537$
(from Set~AT3)
with imposed field $\Bex/\Beq \approx 4.3$.
Time is normalized by the turnover time 
$(\urms \kf)^{-1}$.
Dashed lines: averages over the resetting intervals.
We take the average of many ($\ge 10$) such intervals.
}\label{fig:reset}
\end{figure}

\subsubsection{$\Rm$ dependence}

To study the influence of $\Rm$, we performed simulations
with the higher value $18.2$;
see Set~AT2 in Table~1.
Figure~\ref{fig:AT2} shows that for this set
the quenching of 
$\eta_\perp$ and $\eta_\parallel$ is reduced mildly.
The $\mu$ increases more rapidly with $\Bex$ compared to Set~AT1 and seems to saturate at large fields.
Moreover, we have performed simulations with a fixed value of
$\Bex/\Beq \approx 4.3$, but $\Rm$ increasing from 0.07 to 537;
see Figure~\ref{fig:rm}.
For the largest values of $\Rm$, the resetting 
of the test solutions (see \Sec{sec:testfield})
is most critical,
but it turns out that the resulting values of $\eta_\perp$ and
$\eta_\parallel$ show clear plateaus where statistically stable
averages can be taken; see \Fig{fig:reset} for 
an example.

At low $\Rm$ we do not see much anisotropy, but for $\Rm>1$,
$\eta_\perp$ becomes significantly larger than $\eta_\parallel$. 
Interestingly, at about $\Rm=10$, $\eta_\parallel$ reaches a maximum,
whereas $\eta_\perp$ increases even at the largest $\Rm$, as does $\mu$.
We find again that $\eta_\perp-\mu/2$ is almost identical to $\eta_\parallel$.

It has been reported earlier that in forced hydrodynamic turbulence $\etat$ increases linearly 
with $\Rm$ at smaller values and saturates beyond $\Rm \approx10$ \citep{SBS08}. 
However this is not so in our hydromagnetic turbulence.
Unfortunately, the instability of the test problems for high $\Rm$
prevents us from looking further for a possible saturation.

\subsubsection{Incoherent $\alpha$ effect}
\label{sec:incoh}

\begin{figure}[t]
\centering
\includegraphics[width=0.80\columnwidth]{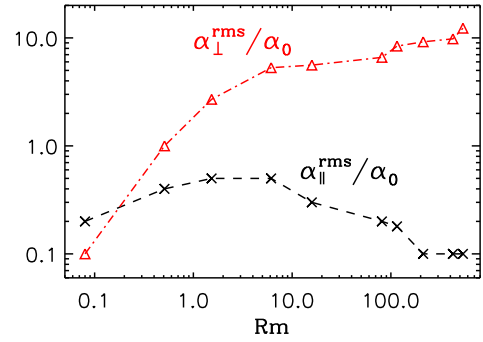}
\caption{As \Fig{fig:rm}, but showing the fluctuations of $\alpha$
as functions of $\Rm$. Crosses: $\alpha_\parallel^{\rm rms}$; 
triangles: $\alpha_\perp^{\rm rms}$. 
}\label{fig:alfluc}
\end{figure}

For non-helical isotropic forcing, the $\alpha$ tensor vanishes on average   
when rotation or stratification is absent.
As emphasized by \cite{BRRK08}, however,
its fluctuations, also referred to as `incoherent $\alpha$ effect',
may in general have relevance for dynamo processes, 
especially if they interact with large-scale shear \citep{VB97,HMWS11,MB12}.
In our simulations they are too weak to lead to self-excitation though.
In \Fig{fig:alfluc} we show the
volume averaged temporal
fluctuations of $\alpha_\perp$ and $\alpha_\parallel$ as functions of $\Rm$
in terms of their rms values, defined as
$\bra{\alpha_\perp^2}_t^{1/2}$ and $\bra{\alpha_\|^2}_t^{1/2}$, respectively, 
where the subscript $t$ refers to time averaging.
While $\alpha_\perp^{\rm rms}$ increases with $\Rm$, $\alpha_\|^{\rm rms}$
increases only slightly at moderate $\Rm$, but decreases beyond $\Rm\approx5$.
Fluctuations in $z$ could also be important and would increase the rms
values of $\alpha_\perp$ and $\alpha_\|$, but have here been ignored.

\subsection{Stratified Convection}

Finally we turn to convection, in which
already in the absence of a magnetic field a preferred direction is set 
by gravity and thus density stratification.
All the relevant transport coefficients are measured
using \tfz\ with wavenumber $k=k_1$, except that in one case
we also consider $k=0$.
As in the case of homogeneous forced turbulence,
we present time-averaged results, but owing to the intrinsic
inhomogeneity of the setup, no $z$ averaging is performed by default.
Error bars are generated as described for forced turbulence.

In deriving quenching characteristics for an inhomogeneous turbulence
from numerical experiments with an imposed (uniform) field, one has to
remember that the actually quenching mean field needs not coincide with
the imposed one. In general, as a consequence of \Eq{meanemf1}, a mean
electromotive force is caused by $\BBex$ which in turn can give rise to
an additional constituent of $\meanBB$.
This could of course not happen in our setups with forcing, 
as there the generated $\meanEMF$ is uniform.
For convection, however, the transport coefficients are at least $z$
dependent (for \tfz) and the $x$ and $y$ components of $\meanEMF$ will
result in $\meanBB\ne\BBex$ with both a modification of the imposed
component and the generation of one or even two components orthogonal to it.
This applies for horizontal as well as vertical $\BBex$.

\subsubsection{Non-rotating convection}

\begin{figure}[t]
\centering
\includegraphics[width=0.48\textwidth]{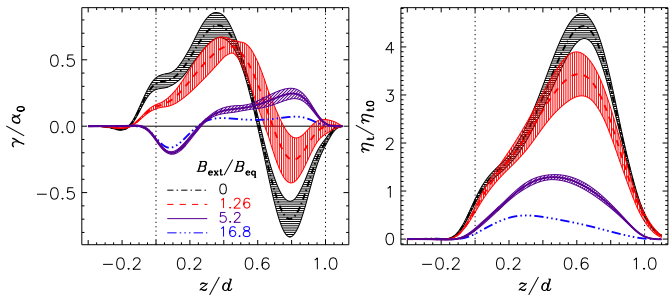}
\caption{
Dependences of $\gamma$ (left) and $\etat$ (right) on the vertical coordinate $z$ for  different $\Bex/\Beq$ (Set~CR0).
Hatched areas: errors
(not shown for $\Bex/\Beq=16.8$ as indistinguishable from the mean).
Dotted lines at $z=0,1$: boundaries of the convectively unstable region.
}\label{fig:CR0prof}
\end{figure}

First we present results for
the simplest situation without rotation or
large-scale shear
(Set~CR0, listed in Table~1). 
No 
(coherent)
$\alpha$ effect is expected, but turbulent pumping, i.e., a $\gamma$ effect, should occur due to
the inhomogeneities caused by stratification and boundaries.
Figure~\ref{fig:CR0prof} presents  profiles of 
$\gamma$ and $\etat$ for four different values of the imposed magnetic field $\Bex$
from zero to $\approx 16.8 \Beq$. We see that the unquenched
profiles of $\gamma$ and $\etat$ are similar to what has been found by \citet{KKB09a}
and that even when $\Bex/\Beq\gtrsim 1$, at least $\etat$ is not quenched much.
However, for $\Bex/\Beq > 2$, both $\gamma$ and $\etat$ are suppressed 
significantly, and $\gamma$ is even changing sign. 
Moreover, the level of fluctuations is markedly
reduced at the highest $\Bex/\Beq$
and the convection itself is suppressed to the extent that it only shows
elongated cells; see \Tab{tab:runs} for the reduction of $\urms$ (cf.\ $\Rm$).
This is a consequence of our choice of using relatively small values
of $\Rm$ and $\Rey$.
As a consequence, the convection is only mildly supercritical and
therefore more vulnerable to quenching.

For weak and moderately strong fields, negative (positive) values of $\gamma$
are seen in the upper (lower) part of the domain, which corresponds to downward (upward)
pumping, i.e., toward the middle of the convection zone.
These directions are
just opposite to what analytic theory predicts for {\em uniform}
mean fields, namely 
that the pumping is directed
away from the maximum of the turbulence intensity.
The obtained behavior agrees, however, 
with the findings of \citet{KKB09a} for harmonic test fields with $k=k_1$
which are also employed in this section. 
For stronger fields the sign is reversed, as expected for
magnetic buoyancy \citep{KP93}.
In \Sec{sec:pumpuni} we will show
results for uniform test fields ($k=0$) and compare them with the theoretical prediction.

\begin{figure}[t]
\centering
\includegraphics[width=0.4\textwidth]{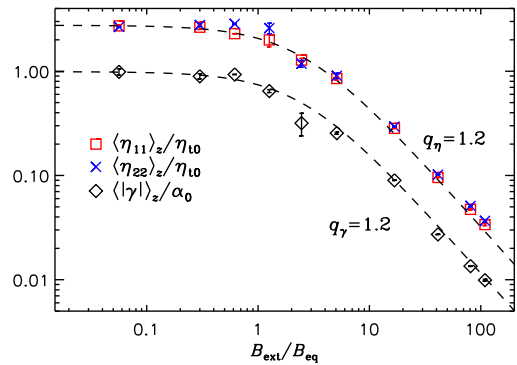}
\caption{Results from Set~CR0: Dependences of $\bra{\eta_{11}}_z$ 
(squares) $\bra{\eta_{22}}_z$ 
(crosses) and $\bra{|\gamma|}_z$ (diamonds) on $\Bex/\Beq$.
Dashed lines: fits from formula~\eqref{scalling} with
exponents $q_{\eta,\gamma}=1.2$.}
\label{fig:CR0}
\end{figure}

\begin{figure}[!h]
\centering
\hspace*{-1mm}\includegraphics[width=0.495\textwidth]{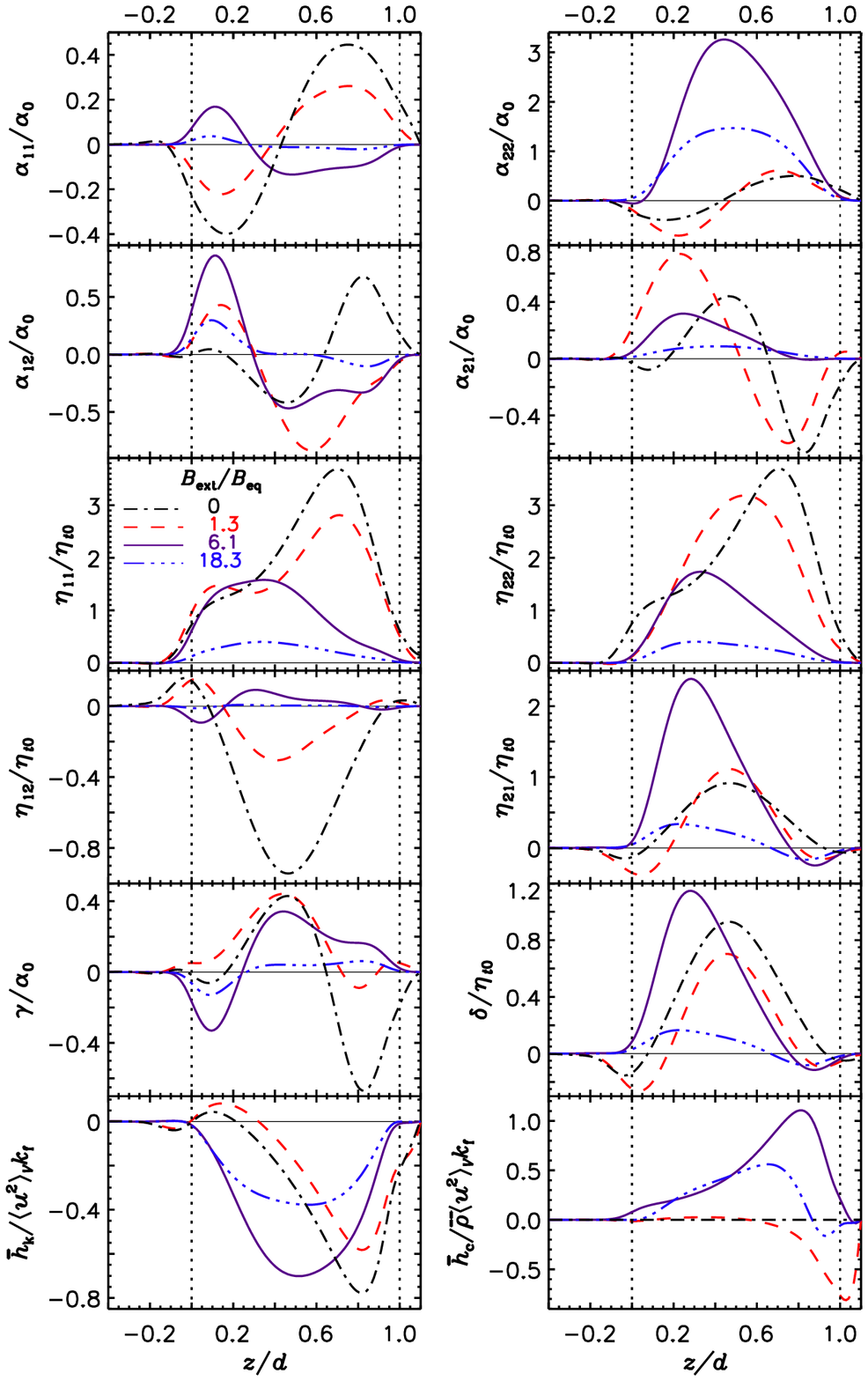}
\caption{Results from Set CR1:
Variations of $\alpha_{ij}$, $\eta_{ij}$, $\gamma$, $\delta$,
and the mean kinetic and current helicity,
${\overline h}_{\rm k}$ and ${\overline h}_{\rm c}$, respectively,
along the vertical coordinate $z$ at different field strengths.
${\overline h}_{\rm k}$ and ${\overline h}_{\rm c}$ are normalized by
$\bra{\uu^2}_{V}\kf$ and $\overline{\rho}\bra{\uu^2}_{V}\kf$, respectively,
where $\bra{\cdot}_V$ indicates volume averaging.
Dotted lines at $z/d=0,1$: boundaries of the convectively unstable region.
}\label{fig:calpha}
\end{figure}

The coefficients are intrinsically $z$ dependent, but
for the sake of clarity in presenting their dependences on $\Bex$,
we calculate the averages of $\eta_{11,22}$
and $|\gamma|$ over a certain $z$ extent,
typically $0.2 \le z/d \le 0.9$. Other intervals or even the degenerate case of fixed values of $z$, however,
yield very similar quenching behaviors.
In Figure~\ref{fig:CR0}, we present $\eta_{11,22}$ and $\gamma$,
averaged in this way, in dependence on
$\Bex$.
Fitting the data with the formula~\eqref{scalling} we find
$q_{\gamma,\eta}=1.2$ which is very close to
our earlier results for the Roberts flow, 
but slightly larger than those found in forced turbulence.
Finding the same quenching dependence of $\gamma$ and $\etat$
seems sensible in the light of the result of the linear theory of \cite{RS75}, 
$\gamma=-\partial_z \etat/2$.
When we normalize $\Bex$ with $\Beqz$ we find for the exponents $q_{\gamma,\eta}\approx2.2$.
The value of $q_\gamma$ disagrees with the analytical result of \cite{RK06}.
However, this was derived for turbulent pumping being 
caused by the density gradient
which 
can hardly be dominant
here because of weak density stratification. Therefore our
result is 
closer to the exponent 2 which is found when 
pumping
is caused by 
a gradient
in the turbulence intensity 
instead
(see \Sec{sec:pumpuni} for the validation).

\subsubsection{Rotating convection}

Next we consider rotating convection with the rotation axis aligned
along the $z$ direction ($\theta=0$),
whereas the magnetic field is along the $x$ direction.
We expect an $\alpha$-effect because $\bm{g} \cdot \bm{\varOmega} \ne 0$.
Figure~\ref{fig:calpha} shows 
the profiles of the measured transport coefficients 
at different strengths of the external field
with $\alpha_{ij}$ normalized to the isotropic value for maximum helicity $\alpha_0 = \urms/3$
and $\eta_{ij}$ normalized to $\etatz$.
The main diagonal elements of both tensors are for $\Bex\ne0$ not equal 
because 
the external field is applied along the $x$ direction.
For vanishing and weak $\Bex$,
both $\alpha_{11}$ and $\alpha_{22}$ change sign,
albeit not at exactly the same position;
they are then positive in (roughly) the upper half of the convective zone
and negative in the lower one, 
again consistent with earlier findings of \cite{OSB01} and \cite{KKB09a}. 
Importantly, $\alpha_{11}$ decreases rapidly with increasing $\Bex$.
However, $\alpha_{22}$ increases at first and only later decreases.

\begin{figure}[!h]
\centering
\hspace*{2mm}\includegraphics[width=0.485\textwidth]{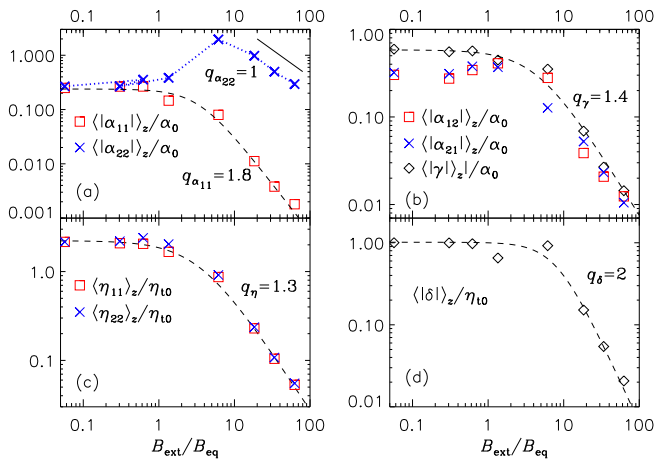}
\caption{Results from Set CR1: Variations of
 $\bra{|\alpha_{11,22}|}_z$ (a), $\bra{|\alpha_{12,21}|}_z$, $\bra{|\gamma]}_z$ (b),
$\bra{\eta_{11,22}}_z$ (c), and $\bra{|\delta|}_z$ (d)
with the external field.
Dashed lines: fits from \eqref{scalling}.}
\label{fig:CR1}
\end{figure}

The components $\eta_{11,22}$ have very similar profiles 
not only
for vanishing, 
but also for
 very strong magnetic field,
differing a bit more for intermediate field strengths.
The off-diagonal components of the $\eta$ tensor
are here of interest mainly in the combination $\delta=(\eta_{21}-\eta_{12})/2$ which characterizes
the $\bm\varOmega\times\JJ$ effect.
In agreement with earlier work for rotating convection \citep{KKB09a},
the sign of $\delta$ is mainly positive, while for rotating forced turbulence,
\cite{BRRK08,BRK12} found it to be negative.
It is also remarkable that $\delta$ is 
only mildly
quenched unless the
magnetic field becomes very strong.
We also see that $\eta_{21}+\eta_{12}$ is not small.
This quantity would vanish in the absence of a magnetic field,
but it is apparently quite sensitive even to weak 
fields.

The lowermost panels in \Fig{fig:calpha} show the
mean kinetic and current helicity as defined by
${\overline{h}}_{\rm k}=\overline{\oo\cdot\uu}$
with $\oo=\nab\times\uu$, and ${\overline{h}}_{\rm c}=
\overline{\jj\cdot\bb}$, respectively.
For weak fields, the kinetic helicity is positive in the lower third
and negative in the upper two thirds of the unstable layer,
while the current helicity changes from positive to negative only at $z \approx 0.6 d$.   
So the expectation of sign equality of the helicities, nourished by
ideas of $\alpha$ quenching
originating from closure approaches \citep{PFL76,KR82},
is only very roughly met. For strong fields, however,
both helicities show only one sign, opposite to each other,
over almost the entire domain.
The current helicity increases first rapidly with the imposed field, but
for the strongest fields both 
helicities begin to be quenched.

In \Fig{fig:CR1}, showing
the absolute values of the transport coefficients averaged over
 $0.2 \le z/d \le0.9$,  
we find $\bra{|\alpha_{11}|}_z$ being quenched according to \eqref{scalling} with 
$q_{\alpha_{11}}=1.8$.
By contrast, $\bra{|\alpha_{22}|}_z$ is growing until $\Bex/\Beq \approx 6$,
where it reaches roughly eight times its unquenched value,
and is falling then, but with a lower power than $\bra{|\alpha_{11}|}_z$,
namely $q_{\alpha_{22}}\approx 1.08$.
Similarly to the $\alpha$ quenching for Roberts forcing, we find that the
quenching exponents are larger when normalizing by $\Beqz$: about $3$
for $\bra{|\alpha_{11}|}_z$ and asymptotically perhaps about $2$ for $\bra{|\alpha_{22}|}_z$.
The power $3$ agrees with earlier analytic results of
\cite{Mof72}, \cite{Rue74}, and \cite{RK93},
while the power 2 agrees with the exponent found by \cite{RK00},
who all normalized by $\Beqz$.

The quantities $\bra{\etat}_z$ and $\bra{|\gamma|}_z$ show also systematic quenching with exponents 
very similar to those found earlier in non-rotating convection ($q=1.2$) and 
in fact identical to the results for Roberts forcing  ($q=1.3$).  
As we have rotation, another relevant quantity is
$\delta$, defined in \Eq{def:ae}, which is essential
for the $\bm\varOmega\times \JJ$ dynamo
in non-helical turbulence with shear,
cf.\ \Eq{axisymE}.
For a recent application to stellar dynamos see \cite{PS09}.
Figure~\ref{fig:CR1} shows the variation 
of $\bra{|\delta|}_z$
with $\Bex$ and we 
find strong quenching  with $q_\delta=2.0$.

\begin{figure}[t]
\centering
\hspace*{.0mm}\includegraphics[width=0.485\textwidth]{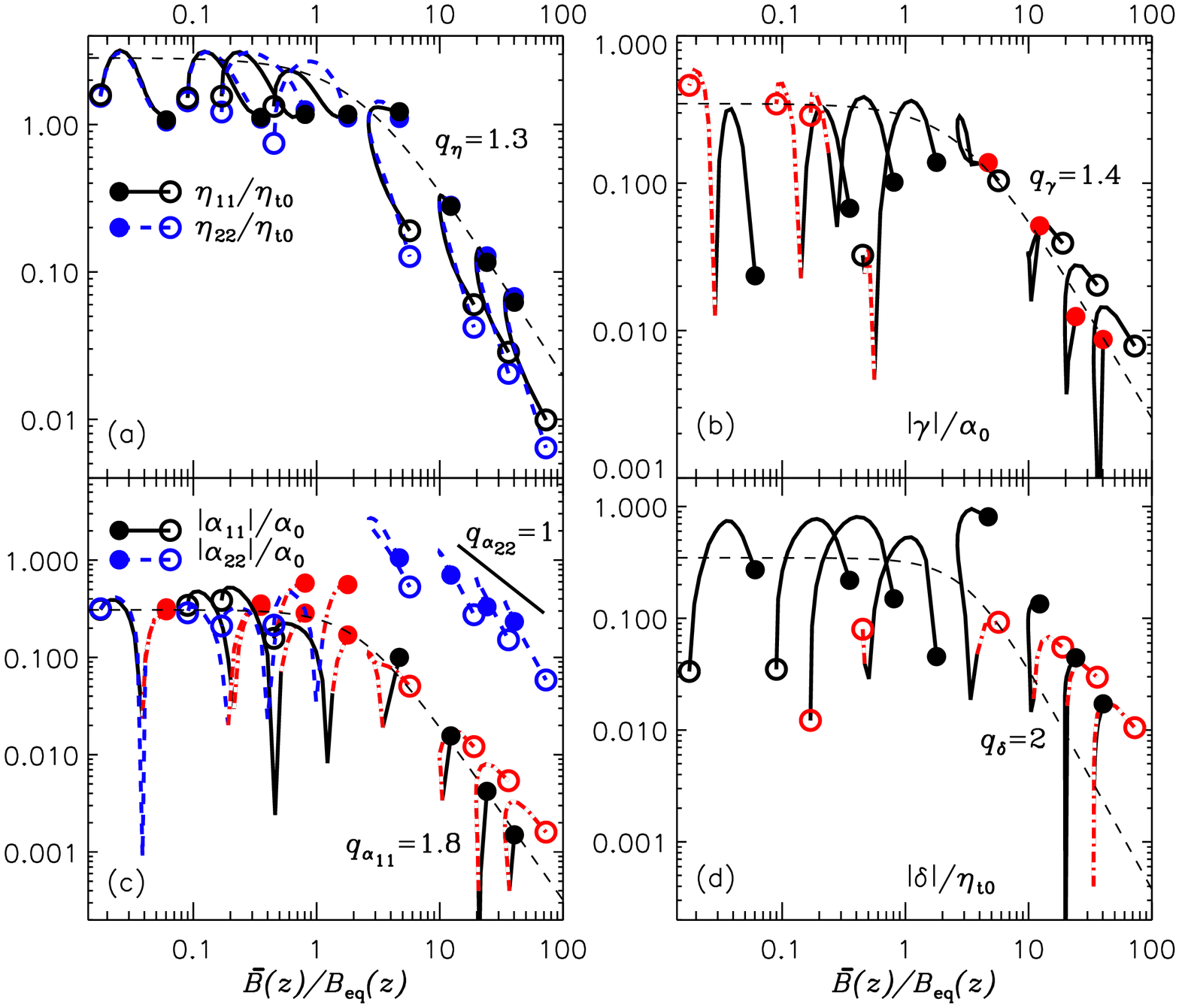}
\caption{As \Fig{fig:CR1} from Set~CR1, but all coefficients are computed locally at 14 
equidistant 
$z$ positions, $ 0.2d \le z \le 0.9d $, in the convective zone
and are plotted against the local value of $\meanB(z)/\Beq(z)$.
Each curve, limited by $\bullet$ and $\circ$, corresponds to a value of $\Bex/\bra{\Beqh(z)}_z 
= 0.21$, 0.83, 2.1, 4.1, 10.3, 20.7, 31.0, and 41.3,
in the order of increasing $\meanB/\Beq$-positions of $\bullet$ (but not necessarily of $\circ$).
$\bullet$ and $\circ$ refer to $z/d=0.2$  and $z/d=0.9$, respectively, thus $z$ is the curve parameter.
Dashed lines: fits to the $z$ averaged quantities from \eqref{scalling}.
Dash-dotted curve sections in (b), (c) and (d) indicate negative $\gamma$, $\alpha_{11,22}$ and $\delta$. 
}
\label{fig:local}
\end{figure}

\begin{figure*}[t]
\centering
\hspace*{-5mm}\includegraphics[width=\linewidth]{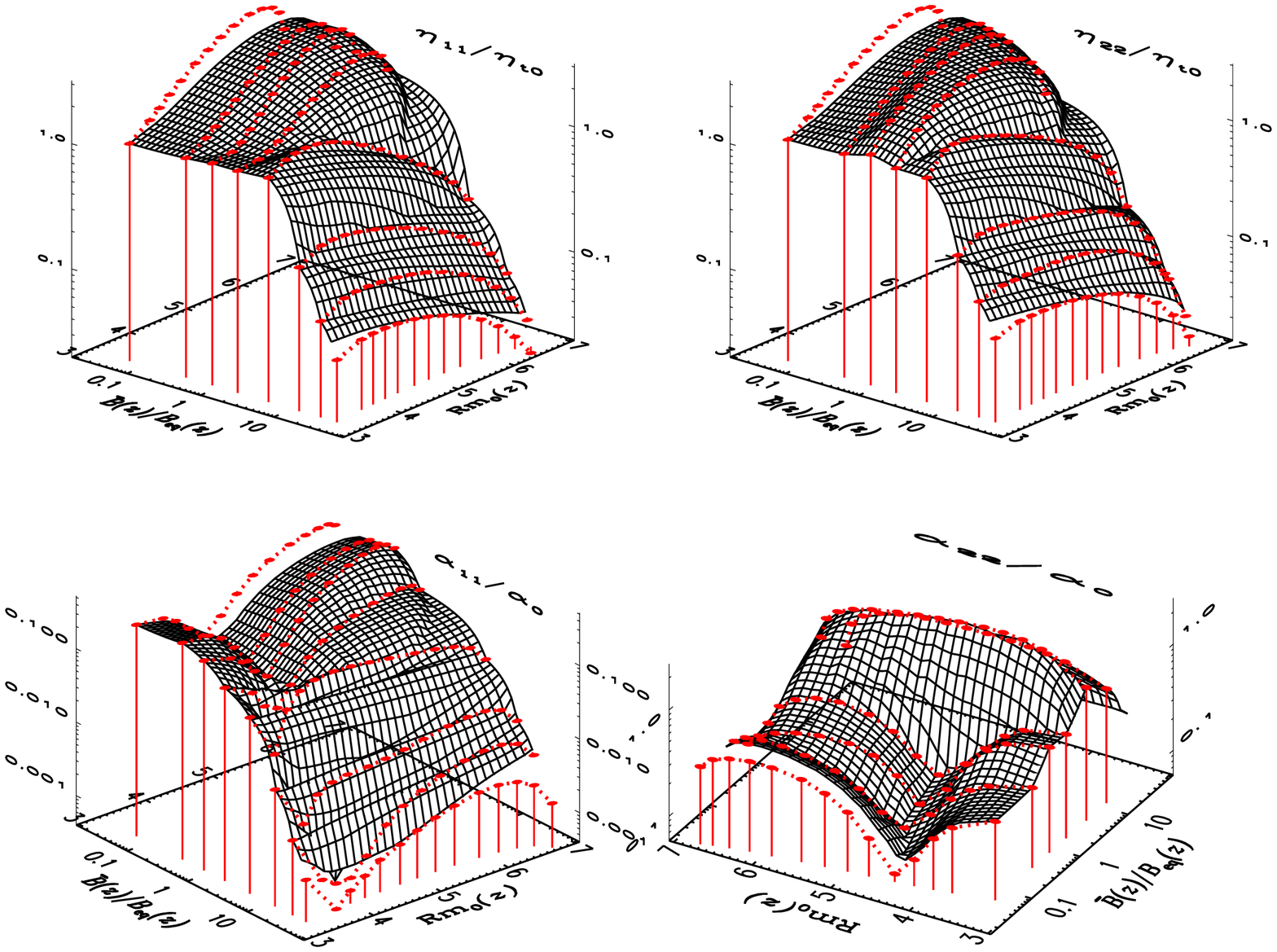}
\caption{As \Fig{fig:local} (Set~CR1), but coefficients shown as functions of $\meanBB/\Beq$ and $\Rmz$,
both depending on $z$.
Upper panels: $\eta_{11}$ and $\eta_{22}$; lower panels: $\alpha_{11}$ and $\alpha_{22}$.
Bullets: data points. Surface: linear interpoland over a triangular grid.
Dotted lines connect points which belong to the same $\Bex/\bra{\Beqh(z)}_z$. }
\label{fig:CR13d}
\end{figure*}

As mentioned above, for the inhomogeneous turbulence in convection
we must take into account that $\meanBB\ne \BBex$.
Therefore we show in \Fig{fig:local}
for Set~CR1 (cf.\ \Fig{fig:CR1}), 
how the transport coefficients are quenched with the local 
$\meanB(z)/\Beq(z)$.
For comparison, the fit to the $z$ averaged quantities from
\eqref{scalling} is shown by the dashed lines.
A more appropriate representation is obtained by considering
the turbulent transport coefficients as functions of both the local $\meanB(z)$ and $\Rmz(z)$,
as they should also depend on the intrinsic (unquenched) local strength of the turbulence.
This view is provided in \Fig{fig:CR13d} where the arguments in
the $(\meanB/\Beq,\Rmz)$ plane were formed by taking both quantities
from the same set of $z$ positions within the convection zone for eight
different values of $\Bex$.
The shown surface was then obtained by linear interpolation
over a Delaunay triangulation of the irregularly spaced arguments.
In $\eta_{11,22}$ we see for fixed $\Rmz$ the common power-law quenching
behavior, while the dependence on $\Rmz$ for fixed $\meanB/\Beq$ grows
until saturation for small $\meanB/\Beq\lesssim 5$, but falling beyond.
$\alpha_{11}$ shows a similar power-law behavior with $\meanB/\Beq$ for
fixed $\Rmz$.
However, the dependence on $\Rmz$ is non-monotonic for $\meanB/\Beq\lesssim 5$
with a minimum between $\Rmz=3$ and 6.
As already indicated by \Fig{fig:CR1}, the behavior of $\alpha_{22}$ is
different in that, it is first growing with $\meanB/\Beq$ reaching a
maximum at $\approx 5$ for all values of $\Rmz$.
As a remarkable feature we see a deeply cut valley with respect to
$\Rmz$ with its floor at $4<\Rmz<5$, ending rather abruptly with a step
at $\meanB/\Beq\gtrsim 3$.
On it and beyond, the $\Rmz$ dependence is weak with a flat maximum.
It remains open, whether the transport coefficients are really local functions
of the two quantities employed, or whether there is also a generic
dependence on the local mean current density.
In addition, non-locality of turbulent transport has been ignored throughout,
which is only permissible at large enough scale separation; see \cite{RB12}.
An attempt to fit to a simple ansatz
$\sim 1/\big(1 + p \Rmz^r (\meanB/\Beq) ^q\big)$ was not satisfactory.
Note that the dependences on $\meanB/\Beq $ and $\Rmz$ were entangled in the result of \cite{BRRS08} as there  $\meanB$, 
being dynamo generated, could not be varied independently of $\Rmz$.

\begin{figure}[t]
\centering
\includegraphics[width=0.35\textwidth]{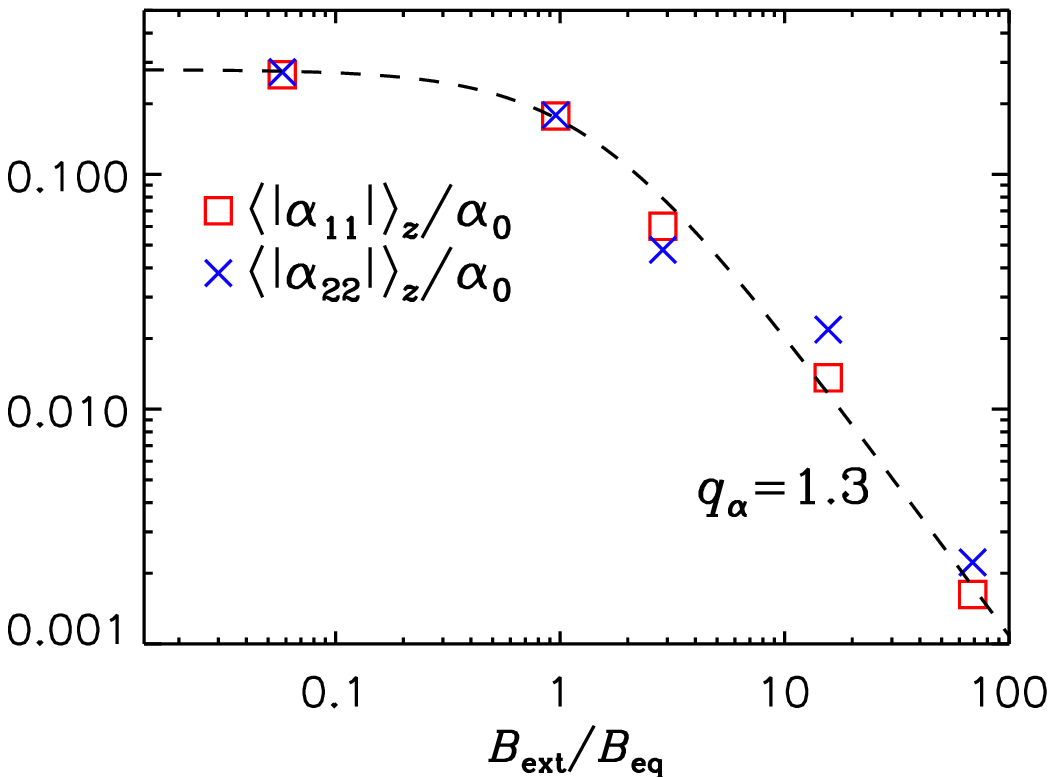}
\caption{Results from Set~CR1Bz
for $\BBex \parallel \zu$: 
Variation of $\bra{|\alpha_{11}|}_z$ (squares) and $\bra{|\alpha_{22}|}_z$ (crosses) with
$\Bex/\Beq$;
cf.\ \Fig{fig:CR1} for Set~CR1.
}
\label{fig:CR1Bzalp}
\end{figure}

In another set of simulations, the external field is applied along 
the vertical direction, see Set~CR1Bz in \Tab{tab:runs}.
Figure~\ref{fig:CR1Bzalp} shows the dependences of $\bra{|\alpha_{11,22}|}_z$
on the external field. 
We see that $\bra{|\alpha_{11}|}_z$ is very close to $\bra{|\alpha_{22}|}_z$,
but now both are quenched with the exponent $q_\alpha=1.3$,
which is smaller than the one of
$\bra{|\alpha_{11}|}_z$ for horizontal external field, see Set~CR1.
Unlike in that case,
$\bra{|\alpha_{22}|}_z$ shows no ``anti-quenching'', cf.\ \Fig{fig:CR1}.
For $\bra{\etat}_z$ and $\bra{|\gamma|}_z$, the quenching exponents
are equal to those
for the horizontal field case, but for $\bra{|\delta|}_z$
we get $q_\delta = 1.8$ instead of 2.
To confirm these results, we have repeated the 
simulations at higher $\Ra$ and $\Rm$ and find similar results;
see Set~CR3Bz in \Tab{tab:runs}.

\begin{figure}[t]
\centering
\includegraphics[width=0.4\textwidth]{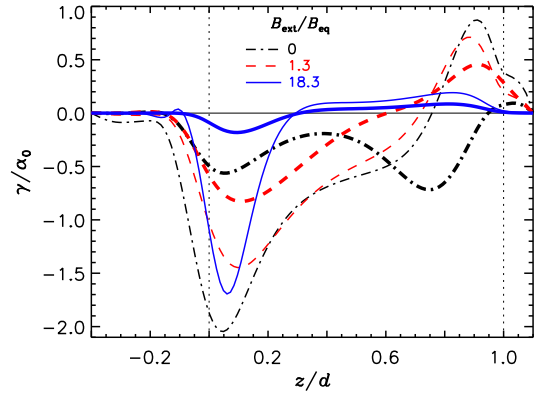}
\caption{Results of Set~CR6: As Set~CR1, but for uniform test fields.
Thick lines: TFZ results for  $\Bex/\Beq = 0$, 1.3 and 18.3;
Thin lines: corresponding profiles of $\gamma_{\rm SOCA}$, see \Eq{eq:gamsoca}.
}
\label{fig:CR3gamma}
\end{figure}

\begin{figure}[t]
\centering
\includegraphics[width=0.493\textwidth]{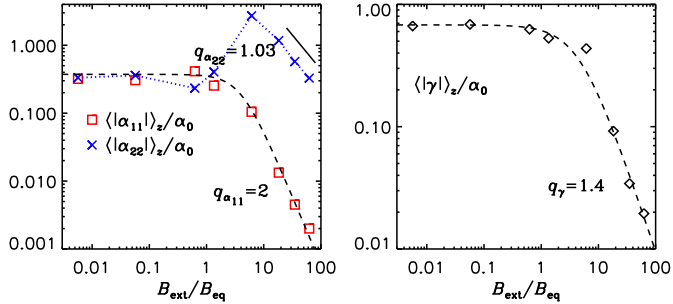}
\caption{Results of Set~CR6: As \Fig{fig:CR1} (Set~CR1), but for uniform test-fields.}
\label{fig:CR3}
\end{figure}

\subsubsection{Turbulent pumping for uniform test fields}
\label{sec:pumpuni}
According to analytic SOCA theory, developed for uniform (or linear)
mean fields, turbulent pumping is related
to the inhomogeneity of the turbulence through \citep{KR80}
\EQ
\gamma_{\rm SOCA} = - (\tau_\gamma/2) \partial_z 
\overline{u_z^2}\, , 
\quad \tau_\gamma\approx \tau_\corr. \label{eq:gamsoca}
\EN
Hence we now employ \tfz\ with {\em uniform} test fields ($k=0$) 
in the sets CR6 and CR7 having horizontal imposed field, see \Tab{tab:runs}.
Figure~\ref{fig:CR3gamma} shows the $z$ profiles of $\gamma$ for different values of $\Bex$
together with those of $\gamma_{\rm SOCA}$
where $\tau_\corr$ has been set to the ($z$ dependent) mixing-length estimate
$ H_p /\bra{\overline{\uu^2}}_{t}^{1/2}$.  
Comparison  reveals that for $\Bex=0$ there is sign agreement 
of $\gamma$ and $\gamma_{\rm SOCA}$ only close to the bottom of the convection zone.
However, already at $\Bex/\Beq=1.3$ the signs match rather well and do so 
perfectly at $\Bex/\Beq=18.3$. The lack of quantitative agreement
in the last two cases can most likely be assigned to the crude estimate of $\tau_\gamma$.  
In the unquenched case, by contrast, no complete agreement can be expected
since SOCA is no longer reliable at $\Rm=\Rmz=3.85$.
At this value we are surely beyond the validity range of SOCA, 
whereas the quenched values of $\Rm$ approach it again. For $\Bex\gg \Beq$, we see that $\gamma$ is strongly quenched,
in particular in the upper two thirds of the convective region.
A comparison with \Fig{fig:CR0prof} 
confirms the strong sensitivity of $\gamma$ -- even in sign -- with
respect to the test-field wavenumber, noticed already in \cite{KKB09a}.
Only for the highest $\Bex/\Beq$ there is some agreement of the $\gamma$ 
profiles for $k=0$ and $k=k_1$.

In Figure~\ref{fig:CR3}, we present the magnetic field dependences of $\bra{|\gamma|}_z$
and $\bra{|\alpha_{11,22}|}_z$, which are similar to what was found 
in Set~CR1 for $z$-dependent test fields, cf.\ \Fig{fig:CR1}.
Set~CR7 with higher $\Rm$ yields essentially the same results.
 
\section{Consequences for mean-field dynamos}

One of the ultimate goals of our work is the application of the numerically obtained
quenching functions to mean-field dynamos that can be validated by comparison
against turbulence simulations and that can 
perhaps eventually be extrapolated to
solar and stellar regimes. One of our striking results
which might have consequences when applied
to mean-field models, is the fact that
for isotropically forced turbulence
the value of $\mu$ keeps increasing with $\Rm$ and exceeds $\etat$ by 
more than a factor of two when $\Bex \gtrsim 10\Beq$,
so turbulent diffusion is enhanced in the direction of the magnetic field
relative to that in the perpendicular direction (cf. \Eq{mu_aniso}), 
but note that $\etat$ is quenched by an order of magnitude and more at such strong $\Bex$.
Furthermore, in \Eq{mu_aniso}, $\epsilon$ is negligible compared with $\etat$,
and so are $\kappa_\perp$ and $\kappa_\parallel$.
Thus the dynamo equation takes the form
\begin{equation}
\partial\meanBB/\partial t=\nab\times(\bm\alpha\cdot\meanBB)+
\etaT\nab^2\meanBB+\mu\nabla^2_\|\meanBB,\label{eq:axmean}
\end{equation}
where an (anisotropic) $\alpha$ effect  has been added.

It is important to note that anisotropic diffusion acts here differently
from what is sometimes assumed in axisymmetric dynamo models \citep{CNC04,JCC07,YNM08,KC11,KC12,KC13}.
To clarify this, let us assume that $\zzz$ is the toroidal direction
and that the toroidal field is dominating, implying $\eee=\zzz$ and $\nabla_\|=0$,
while $x$ and $y$ are coordinates in the meridional plane.
We can then write the magnetic field as
\begin{equation}
\meanBB(x,y,t)=\nab\times(\zzz\meanA_\parallel)+\zzz\meanB_\parallel,  \label{meanBxy}
\end{equation}
and, using this in \Eq{mean_indeq} with \eqref{axisymE}, but $\gamma$, $\delta$, $\kappa_{\|}$ and $\kappa_{\perp}$ neglected,  
 we get \citep[cf.][]{BBMO13}
\begin{alignat}{2}
\partial\meanA_\parallel/\partial t
&= &\,\alpha_A\meanB_\parallel&+\eta_A\nab^2\meanA_\parallel,
\label{EqA}
\\
\partial\meanB_\parallel/\partial t
&= &\,\alpha_B\meanJ_\parallel&+\eta_B\nab^2\meanB_\parallel,
\label{EqB}
\end{alignat}
where
\begin{alignat}{3}
\alpha_A&= \alpha_\parallel,\quad&&&\alpha_B&=\alpha_\perp,\\
\eta_A&=\eta + \eta_\parallel,\quad&&&\eta_B&=\eta+\eta_\perp-\mu/2.
\label{etaAB}
\end{alignat}
With $\eta_\| \approx \eta_\perp - \mu/2$, however, $\eta_A \approx \eta_B$ holds,
hence the diffusion is actually isotropic.
As alluded to above, this is in 
contrast to previously adopted
reasoning by which $\eta_A$ should be much larger than $\eta_B$
\citep[e.g.,][]{CNC04,JCC07,YNM08,KC11}.
We have to stress yet, that the isotropy we found may well be an artefact of the rather specific way of forcing
turbulence by transversal waves.

Furthermore, the amount of quenching assumed in some mean-field models
is rather large, for example \cite{MNM11} employed a reduction of the
magnetic diffusivity by nearly two orders of magnitude
in the lower half of the convection zone
compared to the mixing length estimate.
According to our results,
this would require field strengths that exceed
the equipartition value correspondingly also by two orders of magnitude.
In most of the solar convection zone the equipartition value $\Beqz$
is around 5000 Gauss \citep[see][]{Sti02},
so the mean field strength required for such strong quenching would 
have to reach the unlikely order of several $10^5$ Gauss at the bottom.

Although several mean-field dynamo models 
\cite[e.g.,][]{BMT92,KKT06,GD08,DB12,KN12,PK14} include 
turbulent pumping, its  quenching is usually ignored. 
As an exception, \citet{KKT06} include 
quenching of $\gamma$ with exponent 2
in a formulation with respect to $\Beqz$, which is close to our value of about 2.3. 

The question now is to what extent our new results can be used in modelling
the mean magnetic field evolution either in turbulence simulations of
convectively driven dynamos or even in the Sun.
In recent years, simulations have displayed a wealth of different
behaviors that are hard to explain with our current knowledge.
Examples include the equatorward migration in the simulations of
\cite{K12}, which is only found in the saturated regime of the dynamo
and could therefore be connected with quenching, but in ways that are
even qualitatively unclear.
There are also aspects that might not be possible to capture within
the framework of Cartesian geometry such as the extreme concentration
of toroidal flux belts or wreaths \citep{BBBMT10}, possibly
connected with the dramatic concentration of kinetic helicity toward
low latitudes and near the surface; see Figure~1(b) of \cite{K12}.
One must therefore wait until proper test-field results for azimuthally
averaged fields in spherical shells become available.

With these qualifications in mind, we have to content ourselves with
statements that we can hope are robust.
An example is our finding that the quenching exponents are
of the order of unity and the prefactors typically below unity,
which suggests that the quenched turbulent
transport coefficients should not strongly deviate from their kinematic
values if the magnetic field is comparable with the equipartition value.
If our results should be employed in
a mean-field dynamo model for the Sun, 
those obtained
from convection simulations are the most relevant ones.
Therefore, when restricting oneself to the
coefficients $\alpha$, $\etat$, and $\gamma$,
one could think about choosing
the quenching exponents 3, 2.3, and 2.3, respectively,
in expressions of the form \eqref{scalling},
providing a somewhat stronger quenching than obtained with the usually adopted exponent 2.
However, given that dynamo fields are non-uniform, more elaborated models for the dependence of the transport coefficients
on both the local $\meanB$ and the local $\Beqz$, perhaps even also including a dependence on the local $\meanJ$,
need to be developed. 

\section{Conclusions}

We have measured the quenching of the
turbulent transport coefficients appearing
in the mean-field dynamo equation, in particular 
$\alpha_{ij}$, $\gamma$, $\eta_{ij}$, $\delta$, and $\mu$, by test-field methods.
For this, we have considered three different background flows
on which uniform external magnetic fields with various directions were imposed.
This is of course quite different from the
real situation where quenching occurs due to dynamo-generated
mean fields; see \cite{BRRS08} for a measurement of $\alpha$ and
$\etat$ at large values of $\Rm$  in such a case.
Another aspect to keep in mind is that the magnetic and fluid Reynolds
numbers of our simulations are far too small in comparison with
astrophysical situations.
Extrapolation to $\Rm \to \infty$ is feasible once an asymptotic regime is detected, but
we emphasize that, in agreement with the results of \cite{BRRS08}, 
our maximum value of $\Rm\lesssim 600$ is not yet sufficient.
Nevertheless, the obtained results indicate clear trends that may
well apply to more realistic settings and parameter regimes.

In the setup with Roberts~I forcing, we have found
as a striking property of the quenching behavior
its dependence on whether one normalizes the external
field with the actual or the original
(unquenched) value of the equipartition field
strength, $\Beq$ or $\Beqz$, respectively.
In the former case, the quenching exponent for turbulent diffusivity
and $\alpha$ effect is significantly smaller and closer
 to that found for forced turbulence
and convection (around 1.3). In the latter case, on the other hand, we recover the exponent $4$,
found earlier for $\alpha$ quenching in the Roberts~I flow \citep{RB10}.
Somewhat surprisingly, we find the quenched $\alpha_{ij}$ and $\eta_{ij}$ to be still isotropic in the $xy$ plane,
in contrast to that paper.
However, it is now clear that this is a consequence of their use
of a simplified momentum equation and that the obtained isotropy is physically sensible.

For isotropically forced turbulence, the differences between the two 
normalizations of $\Bex$
are not so large and the exponent based on $\Beqz$ is only
around $1.5$, which is higher than 
what has been found analytically in
\citet{KPR94} and \citet{RK00},
while the exponent based on $\Beq$ is around $1.1$, which
is similar to \citet{G13}.

Finally we have considered rotating stratified convection. Along with
$\alpha$ and $\etat$, we have studied turbulent pumping ($\gamma$) and 
the $\bm\varOmega\times\JJ$ effect ($\delta$).
We find that $\etat$ and $\gamma$ show similar
quenching dependences on $\Bex/\Beq$
(with quenching exponent $q\approx1.3$), while $q$ is about 2 for $\alpha$ and $\delta$.
However, when $\Bex$ is normalized with $\Beqh$ the exponent becomes $3$,
which is agreement with \citet{RK93}.
In non-rotating convection, the quenching of $\gamma$ and $\etat$ is
only slightly weaker compared to the rotating case. 

We have not studied the 
simultaneous
quenching of turbulent transport 
by magnetic field and
rotation
which is particularly important in rapidly rotating stars.
Furthermore, we have not yet applied \tfa\ for convection,
which is a subject of our ongoing work.
It is unclear, however, how useful it would be to consider quenching
with this method, because only one preferred direction is possible.
More general methods would be needed in the
presence of a strong horizontal magnetic field.

\acknowledgments
We thank an anonymous referee for suggestions which improved the presentation.
This work was supported in part by
the European Research Council under the AstroDyn Research Project No.\ 227952,
and the Swedish Research Council Grants No.\ 621-2011-5076 and 2012-5797,
as well as the Research Council of Norway under the FRINATEK grant 231444.
We acknowledge the allocation of computing resources provided by the
Swedish National Allocations Committee at the Center for
Parallel Computers at the Royal Institute of Technology in
Stockholm and the National Supercomputer Centers in Link\"oping,
the High Performance Computing Center North in Ume\aa,
and the Nordic High Performance Computing Center in Reykjavik.

\appendix
\section{Quenching for a single wave flow}
\label{app1}
Assume the forcing in \Eq{eq:vel} to be a single transverse (frozen) wave with time-dependent amplitude, 
\EQ
    \ff = \hat{f}(t) \, \hat{\aaaa}\cos\psi, \quad \psi = \kk\cdot\xx + \phi, \quad \hat{\aaaa}=\frac{\kk \times\ee}{|\kk \times\ee|}, \quad \ee\not\,\parallel \kk \;-\;\text{an arbitrary vector}.
\EN
Then, in the absence of $\BBex$
\begin{align}
    \uu^{(0)} =    \hat{\aaaa} \cos \psi \int_{-\infty}^t \exp \big(\nu k^2(t'-t) \big) \hat{f}(t') dt' =  \hat{u}^{(0)}(t) \,\hat{\aaaa} \cos \psi, \quad 
        p^{(0)} &= \text{const}, \quad \rho^{(0)} = \text{const},
\end{align}
is an exact solution of \Eqs{eq:con}{eq:vel} for arbitrary $\Rey$ as, being a transverse wave, it obeys  $\uu^{(0)}  \cdot\nab  \uu^{(0)} =\bm 0$, $\nab\cdot  \uu^{(0)} =0$ and $\nab^2 \uu^{(0)}  = -k^2 \uu^{(0)}$.
Further 
\EQ
    \bb^{(1)} =  -\Bex k_x  \,\hat{\aaaa}\sin \psi \int_{-\infty}^t \exp \big(\eta k^2(t'-t) \big) \hat{u}^{(0)}(t') dt' = - \hat{\beta}^{(1)}(t) \Bex k_x \, \hat{\aaaa} \sin \psi 
\EN
is an exact solution of the first-order induction equation with horizontal $\BBex$ for arbitrary $\Rm$, again as a consequence of transversality and solenoidality of $\uu^{(0)}$.
Likewise
\EQ
    \uu^{(2)} \approx  -\Bex^2 k_x^2  \,\hat{\aaaa} \cos \psi \int_{-\infty}^t \exp \big(\nu k^2(t'-t) \big) \hat{\beta}^{(1)}(t') dt' /\rho^{(0)}, \quad p^{(2)} \approx -{\bb^{(1)}}^2/2 - \bb^{(1)}\cdot\BBex,
\EN
is an approximate solution of the second order momentum equation as long as $p^{(2)} \ll p^{(0)}$, thus $\rho\approx\rho^{(0)}$. 
As $\uu^{(2)} \sim \uu^{(0)}$ (with a time-dependent factor) the
argument continues to hold in arbitrary orders in $\Bex$. 
$\uu^{(2)} \sim \uu^{(0)}$ would even hold for the less restricting condition
that only the non-constant part of $p^{(2)}$ needs to be negligible.
We conclude that the imposed field does not change the geometry of the flow, but only its amplitude. 
Considering an ensemble of such flows with randomly chosen $\kk$ and $\ee$, the transport coefficients obtained by ensemble averaging would have to be isotropic, even when quenched,
as long as the density remains close to uniform.
\newcommand{\yastroph}[2]{ #1, astro-ph/#2}
\newcommand{\ycsf}[3]{ #1, {Chaos, Solitons \& Fractals,} {#2}, #3}
\newcommand{\yepl}[3]{ #1, {Europhys.\ Lett.,} {#2}, #3}
\newcommand{\yaj}[3]{ #1, {AJ,} {#2}, #3}
\newcommand{\yjgr}[3]{ #1, {J.\ Geophys.\ Res.,} {#2}, #3}
\newcommand{\ysol}[3]{ #1, {Sol.\ Phys.,} {#2}, #3}
\newcommand{\yapj}[3]{ #1, {ApJ,} {#2}, #3}
\newcommand{\ypasp}[3]{ #1, {PASP,} {#2}, #3}
\newcommand{\yapjl}[3]{ #1, {ApJ,} {#2}, #3}
\newcommand{\yapjs}[3]{ #1, {ApJS,} {#2}, #3}
\newcommand{\yija}[3]{ #1, {Int.\ J.\ Astrobiol.,} {#2}, #3}
\newcommand{\yan}[3]{ #1, {Astron.\ Nachr.,} {#2}, #3}
\newcommand{\yzfa}[3]{ #1, {Z.\ f.\ Ap.,} {#2}, #3}
\newcommand{\ymhdn}[3]{ #1, {Magnetohydrodyn.} {#2}, #3}
\newcommand{\yana}[3]{ #1, {A\&A,} {#2}, #3}
\newcommand{\yanas}[3]{ #1, {A\&AS,} {#2}, #3}
\newcommand{\yanar}[3]{ #1, {A\&A Rev.,} {#2}, #3}
\newcommand{\yass}[3]{ #1, {Ap\&SS,} {#2}, #3}
\newcommand{\ygafd}[3]{ #1, {Geophys.\ Astrophys.\ Fluid Dyn.,} {#2}, #3}
\newcommand{\ygrl}[3]{ #1, {Geophys.\ Res.\ Lett.,} {#2}, #3}
\newcommand{\ypasj}[3]{ #1, {Publ.\ Astron.\ Soc.\ Japan,} {#2}, #3}
\newcommand{\yjfm}[3]{ #1, {J.\ Fluid Mech.,} {#2}, #3}
\newcommand{\ypepi}[3]{ #1, {Phys.\ Earth Planet.\ Int.,} {#2}, #3}
\newcommand{\ypf}[3]{ #1, {Phys.\ Fluids,} {#2}, #3}
\newcommand{\ypfb}[3]{ #1, {Phys.\ Fluids B,} {#2}, #3}
\newcommand{\ypp}[3]{ #1, {Phys.\ Plasmas,} {#2}, #3}
\newcommand{\ysov}[3]{ #1, {Sov.\ Astron.,} {#2}, #3}
\newcommand{\ysovl}[3]{ #1, {Sov.\ Astron.\ Lett.,} {#2}, #3}
\newcommand{\yjetp}[3]{ #1, {Sov.\ Phys.\ JETP,} {#2}, #3}
\newcommand{\yphy}[3]{ #1, {Physica,} {#2}, #3}
\newcommand{\yaraa}[3]{ #1, {ARA\&A,} {#2}, #3}
\newcommand{\yanf}[3]{ #1, {Ann. Rev. Fluid Mech.,} {#2}, #3}
\newcommand{\yrpp}[3]{ #1, {Rep.\ Prog.\ Phys.,} {#2}, #3}
\newcommand{\yprs}[3]{ #1, {Proc.\ Roy.\ Soc.\ Lond.,} {#2}, #3}
\newcommand{\yprt}[3]{ #1, {Phys.\ Rep.,} {#2}, #3}
\newcommand{\yprl}[3]{ #1, {Phys.\ Rev.\ Lett.,} {#2}, #3}
\newcommand{\yphl}[3]{ #1, {Phys.\ Lett.,} {#2}, #3}
\newcommand{\yptrs}[3]{ #1, {Phil.\ Trans.\ Roy.\ Soc.,} {#2}, #3}
\newcommand{\ymn}[3]{ #1, {MNRAS,} {#2}, #3}
\newcommand{\ynat}[3]{ #1, {Nature,} {#2}, #3}
\newcommand{\yptrsa}[3]{ #1, {Phil. Trans. Roy. Soc. London A,} {#2}, #3}
\newcommand{\ysci}[3]{ #1, {Science,} {#2}, #3}
\newcommand{\ysph}[3]{ #1, {Solar Phys.,} {#2}, #3}
\newcommand{\ypr}[3]{ #1, {Phys.\ Rev.,} {#2}, #3}
\newcommand{\ypre}[3]{ #1, {Phys.\ Rev.\ E,} {#2}, #3}
\newcommand{\ypnas}[3]{ #1, {Proc.\ Nat.\ Acad.\ Sci.,} {#2}, #3}
\newcommand{\yicarus}[3]{ #1, {Icarus,} {#2}, #3}
\newcommand{\yspd}[3]{ #1, {Sov.\ Phys.\ Dokl.,} {#2}, #3}
\newcommand{\yjcp}[3]{ #1, {J.\ Comput.\ Phys.,} {#2}, #3}
\newcommand{\yjour}[4]{ #1, {#2}, {#3}, #4}
\newcommand{\yprep}[2]{ #1, {\sf #2}}
\newcommand{\ybook}[3]{ #1, {#2} (#3)}
\newcommand{\yproc}[5]{ #1, in {#3}, ed.\ #4 (#5), #2}
\newcommand{\pproc}[4]{ #1, in {#2}, ed.\ #3 (#4), (in press)}
\newcommand{\pprocc}[5]{ #1, in {#2}, ed.\ #3 (#4, #5)}
\newcommand{\pmn}[1]{ #1, {MNRAS}, to be published}
\newcommand{\pana}[1]{ #1, {A\&A}, to be published}
\newcommand{\papj}[1]{ #1, {ApJ}, to be published}
\newcommand{\ppapj}[3]{ #1, {ApJ}, {#2}, to be published in the #3 issue}
\newcommand{\sprl}[1]{ #1, {PRL}, submitted}
\newcommand{\sapj}[1]{ #1, {ApJ}, submitted}
\newcommand{\sana}[1]{ #1, {A\&A}, submitted}
\newcommand{\smn}[1]{ #1, {MNRAS}, submitted}

\vfill\bigskip\noindent\tiny\begin{verbatim}
$Header: /var/cvs/brandenb/tex/karak/aniso/paper.tex,v 1.378 2014/08/20 17:56:08 rei Exp $
\end{verbatim}

\end{document}